\documentclass[journal=langd5]{achemso}
\usepackage[T1]{fontenc} 

\usepackage{amssymb,amsmath}
\usepackage{csquotes}
\usepackage{siunitx}
\usepackage{graphicx}
\usepackage{hyperref}
\usepackage[capitalise]{cleveref}

\usepackage[usenames,dvipsnames]{color}
\usepackage[normalem]{ulem}
\newcommand{\bfuwe}[1]{#1}

\newcommand{\tens}[1]{\boldsymbol{\mathbf{\underline{#1}}}} \newcommand{\xiz}{\xi_{\zeta}}
\newcommand{\xihz}{\xi_{h+\zeta}}
\newcommand{\f}[1]{f_\text{#1}}  \newcommand{\ha}{h_\mathrm{a}}  \newcommand{\thetaeq}{\theta_\text{eq}}  \renewcommand{\vec}[1]{\mathbf{#1}}

\title{Drops on polymer brushes -- advances in thin-film modelling of adaptive substrates}

\author{Simon Hartmann}
\email{s.hartmann@uni-muenster.de}
\affiliation{Institut f\"ur Theoretische Physik, Universit\"at M\"unster, Wilhelm Klemm Str.\ 9, D-48149 M\"unster, Germany}
\affiliation{Center of Nonlinear Science (CeNoS), Universit\"at M\"unster, Corrensstr.\ 2, 48149 M\"unster, Germany}

\author{Jan Diekmann}
\email{jan.diekmann@uni-muenster.de}
\affiliation{Institut f\"ur Theoretische Physik, Universit\"at M\"unster, Wilhelm Klemm Str.\ 9, D-48149 M\"unster, Germany}

\author{Daniel Greve}
\email{daniel.greve@uni-muenster.de}
\affiliation{Institut f\"ur Theoretische Physik, Universit\"at M\"unster, Wilhelm Klemm Str.\ 9, D-48149 M\"unster, Germany}

\author{Uwe Thiele}
\email{u.thiele@uni-muenster.de}
\affiliation{Institut f\"ur Theoretische Physik, Universit\"at M\"unster, Wilhelm Klemm Str.\ 9, D-48149 M\"unster, Germany}
\alsoaffiliation{Center of Nonlinear Science (CeNoS), Universit\"at M\"unster, Corrensstr.\ 2, 48149 M\"unster, Germany}
\abbreviations{\ttjan{ADD}}
\begin{document}
\begin{abstract}
  We briefly review recent advances in the hydrodynamic modeling of the dynamics of droplets on adaptive substrates, \bfuwe{in particular,} solids that are covered by polymer brushes. Thereby, the focus are  long-wave and full-curvature variants of mesoscopic \bfuwe{hydrodynamic models} in gradient dynamics form. After introducing the approach for films/drops of nonvolatile simple liquids on rigid smooth solid substrate, it is first expanded to an arbitrary number of coupled degrees of freedom, before considering the specific case of drops of volatile liquids on brush-covered solids. After presenting the model its usage is illustrated by briefly considering the natural and forced spreading of drops of nonvolatile liquids on a horizontal brush-covered substrate as well as drops sliding down \bfuwe{a brush-covered incline.} Finally, also volatile liquids are considered.
\end{abstract}

\maketitle

\section{Introduction}

The wetting of and dewetting from various substrates by simple and complex liquids frequently occurs in various natural and technological contexts~\cite{GennesBrochardWyartQuere2004,StarovVelardeRadke2007,Bormashenko2017}: Rain drops form intricate patterns on a window pane, spilled coffee spreads forming coffee stains, and badly chosen paint dewets from the surface it is supposed to cover. Modeling the related interface-dominated processes is fundamental to our understanding of liquid behavior. For simple nonvolatile liquids on rigid smooth solid substrates such descriptions are readily available in the form of long-wave (lubrication, thin-film) models, and the dynamic behaviour is well understood~\cite{Genn1985rmp,OrDB1997rmp,BEIM2009rmp,CrMa2009rmp,Wite2020am}. \bfuwe{Often, these thin-film models can alternatively be written in gradient dynamics form~\cite{Mitl1993jcis,Thie2010jpcm} -- a valuable basis for a systematic expansion of the reach of such models \cite{PBMT2005jcp,BCJP2013epje,ThTL2013prl,ThAP2016prf,Thie2018csa}.}

\bfuwe{It is important to recognize that the gradient dynamics approach extends beyond liquids on rigid substrates. Straightforward extensions accommodate scenarios involving flexible and adaptive substrates that attracted much recent interest~\cite{BBSV2018l,AnSn2020arfm}. On the one hand, flexible substrates are characterized by their ability to reversibly change their profile upon deposition of a liquid drop~\cite{PAKL2009sm, StDu2012sm, BCJP2013epje, LWBD2014jfm, AnSn2020arfm}. However, (nearly) no transport of material takes place across the liquid-solid interface. Examples include visco-elastic and elastic soft substrates.}
On the other hand, adaptive substrates \bfuwe{like polymer brushes and hydrogels} take the interaction of liquid and substrate a step further by altering the physico-chemical substrate properties, e.g., their wettability and, potentially, also their topography in response to the presence of liquids~\cite{BBSV2018l}. \bfuwe{In particular, polymer brushes} may swell but also undergo conformational changes when exposed to specific liquids~\cite{HaFK2007sm, CHGM2010nm}. Porous media, with their intricate inner structure, also \bfuwe{represent} adaptive substrates as their wettability changes in dependence of their saturation~\cite{DaHo1999pf,DaHo2000pf,SZKS2003acis,AlRa2004ces,Gamb2014cocis}. Thereby, for all adaptive substrates, liquid may be induced by direct contact, as under a sessile droplet, or it may pass through an external phase as \bfuwe{the ambient atmosphere. Note that also certain types of liquid-infused and slippery substrates may be seen as thin elastic or adaptive substrates \cite{CHRT2023nrc}.}

Our present focus are polymer brushes, i.e.\ polymer chains that are densely grafted to the surface of a rigid solid material~\cite{PLMB2016mme, TGMK2012am, MMMN2003jacs, Doi2013}. The brush can elastically deform and swell due to absorption and imbibition of solvents~\cite{MaMu2005jpcm, LeMu2011jcp, BBSV2018l, MeSB2019m, EDSD2021m} while interacting with the motion of the contact line~\cite{MMHM2011jpcm, WHNK2020l, LSSK2021l}. In contrast to other soft substrates, that often show elastic deformations on large (macroscopic) length scales~\cite{AnSn2020arfm}, the small length of the grafted polymer chains (typically less than \SI{1}{\micro m}) restricts variations of the brush thickness to mesoscopic scales. The swelling is strongly influenced by the mobility of the polymers as well as the miscibility of brush and solvent~\cite{MeSB2019m}. Due to the interplay of the various involved time scales the coupling of substrate and liquid dynamics can result in intricate effects like the stick-slip motion of contact lines in the forced spreading of liquid drops~\cite{WMYT2010scc, SHNF2021acis, LhDa2019csaea}. In consequence, faithful hydrodynamic models for the wetting of adaptive substrates need to incorporate a description of the substrate dynamics.

Here, we review how the gradient dynamics form of mesoscopic hydrodynamics is employed to extend the reach of thin-film models towards brush-covered substrates~\cite{ThHa2020epjt,GrHT2023sm,KHHB2023jcp} and in passing as well show how for volatile liquids the vapor dynamics can be covered~\cite{HDJT2023jfm}. In particular, \bfuwe{the next section} introduces the gradient dynamics approach for films/drops of nonvolatile simple liquids on rigid smooth solid substrates, and then expands it to an arbitrary number of coupled \bfuwe{scalar} degrees of freedom. \bfuwe{The subsequent section} considers the specific case of drops of volatile liquids on brush-covered solids, and \bfuwe{section ``Selected Examples''  illustrates the usage of the presented model by considering} the natural and forced spreading of drops of nonvolatile liquids on a brush, drops sliding down \bfuwe{a brush-covered incline,} and the spreading of a drop of volatile liquid. Finally, \bfuwe{we conclude and give} an outlook.

\section{Gradient dynamics approach to mesoscopic hydrodynamics}
\label{sec:intro-gd}

\subsection{Simple liquid on rigid solid substrate}

For a drop or liquid of nonvolatile partially wetting liquid on a rigid smooth solid substrate the \bfuwe{evolution equation for the film \bfuwe{thickness} profile $h(\vec{x},t)$ can be written in gradient dynamics form as \cite{Mitl1993jcis,Thie2010jpcm}}
\begin{equation}
  \partial_t h \,=\,
  \nabla\cdot\left[Q(h)\nabla\frac{\delta
      \mathcal{F}}{\delta h}\right].
  \label{eq:onefield:gov}
\end{equation}
Here, $Q(h)$ is a \bfuwe{positive} mobility function and $\delta F/\delta h$ denotes the variational derivative of an energy functional
\begin{equation}
  \mathcal{F}[h]\,=\,\int\left[\gamma \xi(|\nabla h|^2) + f(h) \right]\,\mathrm{d}^2r
  \label{eq:en1}
\end{equation}
\bfuwe{that here encompasses interface and wetting energies.} The form of \cref{eq:onefield:gov} is that of a continuity equation, i.e.\ the dynamics conserves the mass \bfuwe{$m=\int h\,\mathrm{d}^2r$.} A nonmass-conserving term proportional to $-\delta F/\delta h$ may be added, e.g., to model evaporation (in so-called one-sided models)~\cite{LyGP2002pre,Thie2010jpcm} or an osmotic influx (in mixture models)~\cite{TJLT2017prl}. \Cref{eq:en1} represents the simple case of a system that is dominated by interfacial effects (capillarity and wettability). Namely, the first term stands for the liquid-gas interface energy with $\gamma$ being the energy density, here, identical to the interface tension, and $\xi(|\nabla h|^2)$ the metric factor. In a full-curvature formulation it is $\xi=\sqrt{1+|\nabla h|^2}$ while a long-wave approximation for shallow drops gives $\xi=1+\frac{1}{2}|\nabla h|^2$ (this is further discussed in section~3 of Ref.~\citenum{Thie2018csa}). The second term stands for the wetting potential that \bfuwe{only} acts on a mesoscopic scale  ($f(h)\to0$ for $h\to\infty$). In the partially wetting case it has a minimum \bfuwe{at the thickness of a nanometric adsorption layer directly related to the} macroscopic equilibrium contact angle. For a more detailed discussion see e.g.\ Ref.~\citenum{Thie2018csa} and references therein.

Introducing the variational derivative of \bfuwe{$\mathcal{F}$ [\cref{eq:en1}]} into \cref{eq:onefield:gov} one obtains the hydrodynamic form of the thin-film equation
\begin{equation}
  \partial_t h \,=\,-\nabla\cdot\left[Q(h)\nabla(\gamma\kappa + \Pi(h))\right]
  \label{eq:film}
\end{equation}
\bfuwe{where} $-\gamma\kappa$ is the Laplace pressure representing capillarity (with $\kappa$ being the curvature), and $\Pi(h)=-df/dh$ is the Derjaguin (or disjoining) pressure. In the case without slip at the substrate the mobility function (as obtained via a long-wave expansion of the Navier-Stokes equations~\cite{OrDB1997rmp}) is $Q(h)=h^3/3\eta$ with $\eta$ being the dynamic viscosity.

Overall, the form~\eqref{eq:onefield:gov} allows one to recognize the thin-film equation as another example of a gradient dynamics of single conserved scalar order parameter fields as, for example, \bfuwe{the Cahn-Hilliard equation that describes the dynamics of phase separation of a binary mixture~\cite{Cahn1965jcp,Thie2010jpcm},} kinetic equations for the evolution of certain surface profiles in epitaxial growth~\cite{GoDN1999pre}, and dynamical density functional theories (DDFT) for the dynamics of colloidal particles~\cite{MaTa1999jcp,ArRa2004jpag}. Note that the form of \bfuwe{such gradient dynamics models with conserved, nonconserved or mixed dynamics} may be derived employing Onsager's variational principle~\cite{Doi2011jpcm} as summarized in the appendix of Ref.~\citenum{Thie2018csa}. Next, we show how the formulation~\eqref{eq:onefield:gov} can be extended to several scalar fields.

\subsection{Case of multiple scalar fields}
In the previous section we have written a simple thin-film model as a mass-conserving gradient dynamics of a single scalar quantity -- the film \bfuwe{thickness} profile, and have mentioned that also nonmass-conserving phenomena may be accommodated. In principle, it should equally be possible to bring models of thin-film flow of complex liquids into such a form if the dynamics is overdamped (Stokes-limit) and there are no persistent fluxes of energy or material in and out of the considered system \cite{Thie2018csa}. Then, the situation is relaxational and the system should approach equilibrium.

\bfuwe{In general,} equation~\eqref{eq:onefield:gov} is extended towards several \bfuwe{equations. If} one considers $n$ scalar quantities $\psi_a(\vec{x},t)$ the resulting $n$ coupled kinetic equations can be written in the form
\begin{equation}
  \partial_t \psi_a=\nabla\cdot \left[\sum_{b=1}^n Q_{ab}
    \nabla \frac{\delta \mathcal{F}}{\delta \psi_b}\right]
  - \sum_{b=1}^n M_{ab} \frac{\delta \mathcal{F}}{\delta \psi_b}
  \label{eq:basics:multi_gradient_dynamics}
\end{equation}
where the $a=1\dots n$. \Cref{eq:basics:multi_gradient_dynamics} \bfuwe{features} mass-conserving and non-mass-conserving terms with $2n^2$ mobility functions \bfuwe{$Q_{ab}$ and $M_{ab}$,} respectively. These form symmetric and positive definite matrices. In the thin-film context the fields may, for instance, represent layer thicknesses, surfactant or solute concentrations, vapor densities or substrate adsorptions. Specific caveats that have to be accounted for are discussed in Ref.~\citenum{ThAP2016prf} for surfactant and solute concentrations, and in Ref.~\citenum{HDJT2023jfm} for vapor densities. Centrally, it is important to chose fields that can be varied independently to obtain valid kinetic equations. For instance, for a thin-film model of a solution one should not chose film \bfuwe{thickness} $h$ and vertically-averaged solute concentration $\phi$, but instead film \bfuwe{thickness} and effective layer \bfuwe{thickness} of solute $\psi=h\phi$.\cite{ThTL2013prl}

In the past years such gradient dynamics models have been proposed for the dewetting of two-layer films \bfuwe{on solid substrates \cite{PBMT2004pre,PBMT2005jcp,BCJP2013epje,JHKP2013sjam,HJKP2015jem,PBJS2018sr},} for the dynamics of liquid films covered by insoluble~\cite{ThAP2012pf} and soluble~\cite{ThAP2016prf} surfactants, dewetting and decomposing films of binary mixtures~\cite{Thie2011epjst,ThTL2013prl}, and the coupled dynamics of shallow drops of volatile liquids and the resulting vapor~\cite{HDJT2023jfm}.  \bfuwe{Although, there, the thermodynamic form sometimes only reformulates models already known in their hydrodynamic form, \cite{WaCM2003jcis,BaGS2005iecr,CrMa2009rmp,NaTh2010n,BMBK2011epje,FrAT2012sm,KaRi2014jfm} it nevertheless allows one to easily thermodynamically validate the models and to identify possible (small and large) thermodynamic inconsistencies and reasons for observed unphysical behavior\footnote{A dramatic example are the ``dewetting waves'' of Ref.~\citenum{FiGo2007jcis} that result from the breaking of Onsager's reciprocity relations, i.e.\ from an early introduction of ``nonreciprocal interactions.''}. Furthermore, models in  gradient dynamics form can be systematically extended by amending the energy functionals, e.g., by incorporating surfactant phase transitions~\cite{ThAP2016prf} or considering full-curvature instead of long-wave variants~\cite{BoTH2018jfm,Thie2018csa}.} Recently the approach has been applied for drops on soft elastic~\cite{HeST2021sm,HEHZ2022prslsapes} and adaptive~\cite{ThHa2020epjt,GrHT2023sm,KHHB2023jcp} substrates. The latter case is reviewed next.

\section{Modeling sessile drops of volatile liquids on polymer brushes}\label{sec:brush-model}
\label{sec:brush-gd}
\subsection{Geometry and fields}
\begin{figure}
  \centering
  \includegraphics[width=\textwidth]{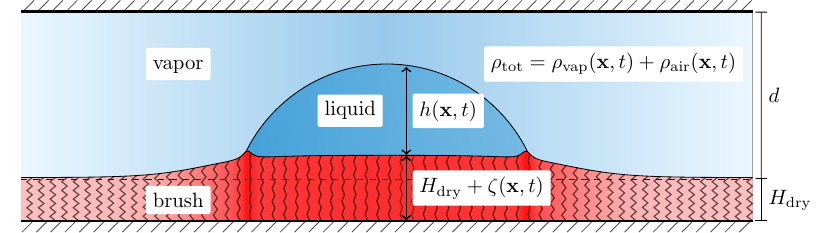}
  \caption{Sketch of a sessile drop of volatile liquid on a polymer brush-covered substrate in a confined geometry formed by two parallel plates separated by a gap of width \(d+H_\mathrm{dry}\). The thickness profile of the liquid film is $h(\vec{x}, t)$ while the brush height profile is $H_\mathrm{dry} + \zeta(\vec{x}, t)$ where $H_\mathrm{dry}$ is the thickness of the dry brush and $\zeta$ is the local effective \bfuwe{thickness} of the imbibed liquid. The gas phase is characterized by the vertically-averaged particle densities of vapor $\rho_\mathrm{vap}(\vec{x}, t)$ and dry air $\rho_\mathrm{air}(\vec{x}, t)$. Note that the sketch is not to scale.}
  \label{fig:brushevap-brush-sketch}
\end{figure}

Next we apply the \bfuwe{introduced general approach} and present a model for drops of volatile liquids on an adaptive substrate, namely, a solid covered by a polymer brush. \bfuwe{It} corresponds to the \bfuwe{model} employed in Ref.~\citenum{KHHB2023jcp} and further builds on models in Refs.~\citenum{ThHa2020epjt,GrHT2023sm,HDJT2023jfm} (that result as limiting cases). In particular, we consider a confined geometry as sketched in \cref{fig:brushevap-brush-sketch}. The employed narrow-gap geometry for the vapor phase, i.e.\ the confinement of the drop in the gap between two parallel plates allows us to describe the system via the coupled dynamics of three fields: the local effective \bfuwe{thickness} of the liquid  imbibed into the brush $\zeta(\vec{x}, t)$, the local \bfuwe{thickness} of the liquid film/drop on top of the brush $h(\vec{x}, t)$, and the vertically-averaged vapor density $\rho_\mathrm{vap}(\vec{x}, t)$. The overdamped dynamics of the system is described by a gradient dynamics as \cref{eq:basics:multi_gradient_dynamics} \bfuwe{with $n=3$.}

To easily keep track of the fluid across the different ``phases'' -- liquid film, brush and vapor -- we express the amount of fluid in the three different regions by the quantities $\psi_1$, $\psi_2$, and $\psi_3$, respectively. Here, for convenience, the three fields all represent particle numbers per substrate area.
Although the usage of the $\psi_a$ makes the mathematical structure more transparent because, e.g., all components of the mobility matrices have the same units, their relation to ``more natural'' fields as, e.g., film \bfuwe{thickness} and vapor density is important for an intuitive understanding of the model. The employed particle numbers per substrate area can be translated into an effective \bfuwe{layer thickness} using the constant particle number density per volume of the liquid $\rho_\mathrm{liq}$ as proportionality factor. In particular, the relation for the local thickness of the liquid film/drop $h(\vec{x}, t)$ is \(\psi_1= \rho_\mathrm{liq}\,h\), for the effective local \bfuwe{thickness} of liquid absorbed by the brush $\zeta(\vec{x}, t)$ it is \(\psi_2 = \rho_\mathrm{liq}\zeta\). This is directly related to the local swelling ratio $\alpha(\vec{x}, t)\ge1$ by $\alpha = (H_\mathrm{dry} + \zeta)/H_\mathrm{dry}$ where \(H_\mathrm{dry}\) stands for the reference thickness of a completely dry brush (\(\alpha = 1\)). A further useful quantity is the local (vertically averaged) part per volume concentration of polymer within the brush $c(\vec{x}, t)=1/\alpha$.

Note that the employed approximation implicitly assumes that the liquid absorption  into the brush is vertically homogeneous. The considered confined geometry also allows us to assume that the vapor distribution in the gap between liquid-vapor interface and upper plate is approximately vertically homogeneous. In consequence, the particle number in the vapor phase per substrate area $\psi_3(\vec{x}, t)$ is related to the vertically averaged vapor particle density \(\rho_\mathrm{vap}(\vec{x}, t)\) via \(\psi_3 = \rho_\mathrm{vap}\,[d-h-\zeta]\). Considering air and vapor as ideal gases, the relative local vapor saturation \(\phi(\vec{x}, t)\) is $\phi = \rho_\mathrm{vap}\,k_B T/p_\mathrm{sat}$ where \(p_\mathrm{sat}\) is \bfuwe{the saturation} pressure of the liquid vapor. For further details of the modeling of the liquid-vapor sub-system see Ref.~\citenum{HDJT2023jfm}.

\subsection{Transport processes}
After having established the governing equations and the fields used in the description, next, we discuss the mobility matrices \(\tens{Q}\) and \(\tens{M}\) in \cref{eq:basics:multi_gradient_dynamics} that are now \(3\times 3\) symmetric positive definite matrices, and reflect all occurring transport processes. The locally mass-conserving lateral transport within the three regions liquid film, brush and vapor are captured by \(\tens{Q}\). Neglecting dynamic coupling between the phases, i.e.\ assuming that no shear stress is transferred across the liquid-vapor and liquid-brush interfaces, the matrix is \bfuwe{diagonal. It accounts for} the viscous motion within the drop with a standard cubic mobility function~\cite{OrDB1997rmp}, diffusive transport \bfuwe{within the brush and within the gas,} both with a standard linear mobility function~\cite{Doi2013}. As a results we have
\begin{equation}
  \tens{Q} = \begin{pmatrix}
    \frac{1}{\rho_\mathrm{liq}}\,\frac{\psi_1^3}{3 \eta} & 0                                      & 0                                    \\
    0                                                    & \frac{1}{k_B T} D_\mathrm{brush} \psi_2 & 0                                    \\
    0                                                    & 0                                      & \frac{1}{k_B T} D_\mathrm{vap} \psi_3
  \end{pmatrix},
  \label{eq:brushevap-conserved-mobility}
\end{equation}
where $\eta$ is the dynamic viscosity of the liquid, and  \(D_\mathrm{vap}\) and \(D_\mathrm{brush}\) are diffusion coefficients of vapor in air and of liquid in the brush, respectively.

All exchange processes between the three \bfuwe{regions are captured} by \(\tens{M}\). As no material leaves or enters the system through the confining plates locally the combined amount of liquid particles across all three regions is conserved. This implies that the sum over columns of \(\tens{M}\) has to be zero. Then, we account for phase change between the liquid film and the vapor (evaporation or condensation) and liquid transfer between the film and the brush layer (imbibition or desiccation) with respective transfer rate constants (or Onsager coefficients) \(M_\mathrm{ev}\) and \(M_\mathrm{im}\). Furthermore, we account for a direct exchange between the brush and the gas phase via a transfer rate constant \(M_\mathrm{ev}'\). In principle, the transfer rates may depend on the state variables \(\psi_a\) -- this option is not followed here. In other words, here, the particle transfer fluxes are proportional to the difference in the chemical potentials between the respective phases. Note that for the condensed phases the assumed constant reference density $\rho_\mathrm{liq}$ implies that the partial pressures are directly proportional to corresponding chemical potentials. As a results we have
\begin{equation}
  \tens{M} = \begin{pmatrix}
    M_\mathrm{im} + M_\mathrm{ev} & -M_\mathrm{im}                 & -M_\mathrm{ev}               \\
    -M_\mathrm{im}                & M_\mathrm{im} + M_\mathrm{ev}' & -M_\mathrm{ev}'              \\
    -M_\mathrm{ev}                & -M_\mathrm{ev}'                & M_\mathrm{ev}+M_\mathrm{ev}'
  \end{pmatrix}.
  \label{eq:brushevap-nonconserved-mobility}
\end{equation}
Notably, the symmetry of the matrix reflects the fact that all transfer processes are allowed to occur in both directions. As stated above, mass conservation across the three regions implies that the dynamics of the the total particle number per area (i.e.\ sum of the three fields) follows a continuity equation \(\partial_t (\psi_1 + \psi_2 + \psi_3) = -\nabla\cdot\vec{j}\) where \(\vec{j}\) is the total flux.

The presented description of the transport processes allows for the transfer of liquid between vapor and brush even when they are separated by a thick film or drop. This is fixed by amending the transfer coefficient \(M_\mathrm{ev}'\) with a smooth step function in such a way that it approaches zero when the liquid \bfuwe{thickness} \(h(\vec{x}, t)\) crosses a small threshold value. Similarly, the transfer coefficients \(M_\mathrm{ev}\), \(M_\mathrm{im}\) are modulated to suppress any transfer of liquid from and to the very thin adsorption layer that covers the brush even far away of a drop. The existence of such a film is a consequence of the employed model for partial wettability (see next section). For details see Ref.~\citenum{KHHB2023jcp}. This approach avoids spurious transport processes.

\subsection{Free energy functional}\label{sec:brushevap-en}
After having introduced the mobilities characterizing the transport and transfer processes, next, we discuss the underlying energy functional \(\mathcal{F}[\psi_1,\psi_2,\psi_3]\) for the combined \bfuwe{liquid} film, brush, and vapor system. Its three variations \(\delta \mathcal{F}/\delta \psi_a\) give the respective local chemical \bfuwe{potentials. Their gradients} and their differences between phases consistently drive all dynamic processes. For convenience, we state the energy in terms of \(h\), \(\zeta \), and \(\rho_\mathrm{vap}\) as
\begin{align}
  \mathcal{F} = \int_\Omega \Bigg[
    \underbrace{\xihz\, \gamma_\mathrm{lg}}_\text{liquid-gas\,interface\,energy}
    + \underbrace{\xiz\, \gamma_\mathrm{bl}(\zeta)}_\text{liquid-brush\,interface\,energy}
    + \underbrace{\xiz f_\mathrm{wet}(h,\,\zeta)}_\text{wetting potential}
    + \underbrace{f_\mathrm{brush}(\zeta)}_\text{brush energy}\nonumber\\
    + \underbrace{(h+\zeta) f_\mathrm{liq}(\rho_\mathrm{liq})}_\text{liquid bulk energy}
    + \underbrace{(d-h-\zeta) f_\mathrm{vap}(\rho_\mathrm{vap})}_\text{vapor energy}
    + \underbrace{(d-h-\zeta) f_\mathrm{air}(\rho_\mathrm{air})}_\text{air energy}
    \Bigg]\,\mathrm{d}^2r,\label{eq:brushevap-free-energy-functional}
\end{align}
i.e.\ to perform the variations, the dependencies of \(h\), \(\zeta \), and \(\rho_\mathrm{vap}\) on the $\psi_a$ have to be taken into account. In \cref{eq:brushevap-free-energy-functional} the quantities \(f_\mathrm{liq}\), \(f_\mathrm{vap}\), and \(f_\mathrm{air}\) are the bulk liquid, vapor, and air free energies per volume. They are multiplied by $h+\zeta$ and $d-h-\zeta$, respectively, to obtain energy densities per substrate area. Here, $f_\mathrm{liq}$ is a constant and the gas energies are discussed further below.
Further, \(\gamma_\mathrm{lg} \) and  \(\gamma_\mathrm{bl}(\zeta)\) are the constant liquid-gas and the brush state-dependent liquid-brush interface energy while \(f_\mathrm{wet}(h,\zeta)\) is the brush state and film \bfuwe{thickness}-dependent wetting energy. Interface and wetting energies are all multiplied by the corresponding metric factors to account for the local arclength of the interface. They are defined by \(\xiz=\sqrt{1+|\nabla\zeta|^2}\) and \(\xihz=\sqrt{1+|\nabla(h+\zeta)|^2}\),  i.e.\ in general, the full-curvature formulation is used. Employing this exact parametrization of the interface instead of the long-wave approximation allows for a recovery of the precise mesoscopic equivalent of the macroscopic Neumann condition valid at three phase contact where the adaptive substrate and the liquid-vapor interface meet (for details see Sec.~3.2 of Ref.~\citenum{GrHT2023sm}).

For the wetting potential we employ a simple combination of short- and long-range power laws
\begin{equation}
  \f{wet}(h,\,\zeta) = A_\mathrm{dry}\,c^a  \, \left(-\frac{1}{2 h^2} + \frac{\ha^3}{5 h^5} \right).\label{eq:brush-f_wet_new}
\end{equation}
where $c=1/\alpha=H_\mathrm{dry}/(H_\mathrm{dry} + \zeta)$ and the prefactor $A_\mathrm{dry}\,c^a$ represents a brush state-dependent Hamaker constant. \bfuwe{In the dry case it is well known how to relate it to the macroscopic equilibrium angle via the mesoscopic Young relation $\cos \theta_\mathrm{eq, dry} = 1 + \f{wet}(\ha,\,0) /\gamma_\mathrm{lg}$. An analogeous relation for other brush states is discussed below in section~``Equilibrium contact angles.''} Although, in the simplest case the power $a=1$ \bfuwe{is} used, the exponent may also be employed to accommodate other experimentally observed behavior close to the wetting transition that here occurs as $c\to0$, i.e.\ $\alpha\to\infty$. Note that $\thetaeq$ does not exactly correspond to the macroscopic equilibrium contact angle as there is a small correction to Young's law due to the different adsorption states underneath and outside of a drop. Further note that one may also let the adsorption layer \bfuwe{thickness} $\ha$ depend on the brush state to account for the fact that short- and long-range contributions to the wetting potential $\f{wet}$ will normally not scale identically with the polymer concentration $c$. However, here, this is not further pursued.

The brush-liquid and brush-vapor interface energies adapt like
\begin{equation}
  \gamma_\mathrm{bl}(\zeta) = \gamma_\mathrm{bl,\,dry}\,c^{\hat{a}}\quad\text{and}\quad\gamma_\mathrm{bg}(\zeta) = (\gamma_\mathrm{bg,\,dry} - \gamma_\mathrm{lg}) \,c^{\tilde{a}} + \gamma_\mathrm{lg},\label{eq:brush-gamma-deps}
\end{equation}
respectively. In other words, we employ power law interpolations between the known limits of dry and completely filled (i.e.\ liquid-like brush) brush. In principle, the powers $a, \hat a$, and $\tilde a$ may all be different. However, for simplicity we use linear dependencies in all \bfuwe{considered examples.}

For the central brush energy we adapt the Alexander-de Gennes mean-field approach~\cite{Alex1977jpp, Genn1991casi}. It rests on the assumptions that the brush density is vertically uniform implying that all polymer chains are uniformly stretched and that all chains end at the same height (the brush thickness). Expressed in our notation we have the expression
\begin{align}
  \f{brush}(\zeta) &= H_\mathrm{dry} \, \rho_\mathrm{brush} \, k_B T \, \left[ \frac{\sigma^2}{2c^2} + (1/c-1) \log (1-c) + \chi (1-c) \right].\label{eq:brush-f_brush}
\end{align}
that combines an elastic contribution (first term, cf.~\citenum{Somm2017m}) and mixing contributions (remaining terms, representing entropic and enthalpic part of a Flory-Huggins-type model~\cite{Flory1953}). Thereby, \(\rho_\mathrm{brush}\) is the (constant) monomer density in the dry brush, \(\sigma\) is the dimensionless grafting density, \(\chi\) is the Flory parameter that quantifies the strength of the polymer-liquid interaction, i.e.\ it controls their miscibility. In particular, positive [negative] $\chi$ favors demixing [mixing]. Interestingly, due to the integration over the brush height the parabolic shape of the interaction term is lost implying that the per substrate area mixing energy has only one minimum with respect to the polymer concentration, i.e.\ the adaptivity of the brush suppresses polymer-liquid phase separation (for details see~\citenum{Hartmann2023Munster}).

Vapor and air are treated as ideal gases, i.e.\ their free energy density  is purely entropic, namely,
\begin{equation}
  f_\mathrm{vap} = k_B T \rho_\mathrm{vap} \left[ \log(\Lambda^3\rho_\mathrm{vap}) - 1 \right]
  ~~\text{and}~~
  f_\mathrm{air} = k_B T \rho_\mathrm{air} \left[ \log(\Lambda^3\rho_\mathrm{air}) - 1 \right],
\end{equation}
where \(\Lambda\) is the \bfuwe{standard thermal de Broglie wavelength}. The density $\rho_\mathrm{air}(\vec{x}, t)$ is related to the vapor density $\rho_\mathrm{vap}(\vec{x}, t)$ via the uniform total density $\rho_\mathrm{tot}$ by \(\rho_\mathrm{air} = \rho_\mathrm{tot}-\rho_\mathrm{vap} \). The density $\rho_\mathrm{tot}$ is uniform as overall pressure gradients equilibrate with the speed of sound, i.e.\ much faster than the considered diffusive processes. For details see Ref.~\citenum{HDJT2023jfm}.

This completes the three-field model describing the dynamics of a drop of volatile liquid on a brush-covered substrate. To be better able to relate the model obtained in its ``thermodynamic form'' as gradient dynamics on the underlying energy functional~\eqref{eq:brushevap-free-energy-functional} to thin-film models in the literature we next discuss its ``hydrodynamic form.''

\subsection{Hydrodynamic form of governing equations}

The variations of the energy functional~\eqref{eq:brushevap-free-energy-functional} with respect to the three fields $\psi_a$ are
\begin{equation}
  \begin{aligned}
    \frac{\delta \mathcal{F}}{\delta \psi_1} &= \frac{1}{\rho_\mathrm{liq}} \frac{\delta \mathcal{F}}{\delta h} = \frac{1}{\rho_\mathrm{liq}} \left[ -\gamma_\mathrm{lg} \frac{\Delta (h+\zeta)}{\xihz^3} + \xiz \partial_h f_\mathrm{wet}(h,\,\zeta) + f_\mathrm{liq} \right],                                         \\
    \frac{\delta \mathcal{F}}{\delta \psi_2} &= \frac{1}{\rho_\mathrm{liq}} \frac{\delta \mathcal{F}}{\delta \zeta} = \frac{1}{\rho_\mathrm{liq}} \bigg[ -\gamma_\mathrm{lg} \frac{\Delta (h+\zeta)}{\xihz^3} - \nabla \left\{ \left[ \gamma_\mathrm{bl}(\zeta) + f_\mathrm{wet}(h,\,\zeta) \right] \frac{\nabla \zeta}{\xiz} \right\}\\
      & \hspace{3em}+ \xiz \partial_\zeta \left[ \gamma_\mathrm{bl}(\zeta) + f_\mathrm{wet}(h,\,\zeta) \right] + \partial_\zeta f_\mathrm{brush}(\zeta) + f_\mathrm{liq} \bigg], \\
    \frac{\delta \mathcal{F}}{\delta \psi_3}   & = k_B T \log \left( \frac{\rho_\mathrm{vap}}{\rho_\mathrm{tot} - \rho_\mathrm{vap}} \right).
  \end{aligned}\label{eq:brushevap-pressures}
\end{equation}
Note that we have simplified the first two equations by assuming that the vapor density $\rho_\mathrm{vap}$ is much smaller than the total gas density $\rho_\mathrm{tot}$, and that $\rho_\mathrm{tot}$ is much smaller than the liquid density $\rho_\mathrm{liq}$.

Introducing the variations into the gradient dynamics~\eqref{eq:basics:multi_gradient_dynamics} and using the more intuitive film \bfuwe{thickness} \(h =\psi_1/\rho_\mathrm{liq}\), effective increase in brush height \(\zeta=\psi_2/\rho_\mathrm{liq}\), and the vapor saturation \(\phi = \rho_\mathrm{vap}/\rho_\mathrm{sat} = p_\mathrm{vap}/p_\mathrm{sat} \approx \psi_3/[(d - h) \rho_\mathrm{sat}]\) results in the evolution equations for effective liquid \bfuwe{thicknesses} in the respective phases\footnote{Here, we have used that the brush is much thinner than the gap width, \(\zeta \ll d\).
}
\begin{equation}
  \begin{aligned}
    \partial_t h           & = -\nabla\cdot \vec j_h- j_\mathrm{ev} - j_\mathrm{im}                                                \\
    \partial_t \zeta      & = -\nabla\cdot \vec j_\zeta + j_\mathrm{im} - j_\mathrm{ev}'\\
    \partial_t [(d-h)\phi] & = -\nabla\cdot \vec j_\rho+ \frac{\rho_\mathrm{liq} k_B T}{p_\mathrm{sat}}(j_\mathrm{ev} + j_\mathrm{ev}')
  \end{aligned}\label{eq:brushevap-dynamic_eqs_final}
\end{equation}
with the conserved fluxes
\begin{equation}
  \begin{aligned}
    \vec j_h & = -\frac{h^3}{3\eta} \, \nabla \frac{\delta \mathcal{F}}{\delta h} \\
    \vec j_\zeta & = -\frac{D_\mathrm{brush} \, \zeta}{\rho_\mathrm{liq} k_B T} \, \nabla \frac{\delta \mathcal{F}}{\delta \zeta} \\
    \vec j_\rho & = -D_\mathrm{vap} (d-h) \nabla \phi \label{eq:brushcons-fluxes}
  \end{aligned}
\end{equation}
and the nonconserved fluxes
\begin{equation}
  \begin{aligned}
    j_\mathrm{ev} &= \frac{M_\mathrm{ev}}{\rho_\mathrm{liq}} \left(\frac{\delta \mathcal{F}}{\delta \psi_1} - \frac{\delta \mathcal{F}}{\delta \psi_3}\right)\\
    j_\mathrm{ev}' &= \frac{M_\mathrm{ev}'}{\rho_\mathrm{liq}} \left(\frac{\delta \mathcal{F}}{\delta \psi_2} - \frac{\delta \mathcal{F}}{\delta \psi_3}\right)\\
    j_\mathrm{im} &= \frac{M_\mathrm{im}}{\rho_\mathrm{liq}} \left(\frac{\delta \mathcal{F}}{\delta \psi_1} - \frac{\delta \mathcal{F}}{\delta \psi_2}\right).\label{eq:brushevap-fluxes}
  \end{aligned}
\end{equation}
Eqs.~\eqref{eq:brushevap-dynamic_eqs_final}, \eqref{eq:brushcons-fluxes} and \eqref{eq:brushevap-fluxes} form the complete dynamical model in hydrodynamic form, i.e.\ the equations for the condensed phases are written in the typical form where pressure gradients drive flows of liquid, i.e.\ the fields and fluxes represent film \bfuwe{thicknesses} (volumes per substrate area) and liquid volumes per time, respectively.

\subsection{Simplifications, limiting cases and extensions}

The \bfuwe{reviewed model} describes the dynamics of a drop of volatile liquid on an adaptive substrate, namely, a thin polymer brush that can exchange \bfuwe{liquid} with the drop. Further, drop and brush can exchange liquid with the vapor phase. The model captures the coupled dynamics of liquid in the drop, brush and vapor as well as the transfer between them. The given formulation \bfuwe{employs} a full-curvature form of the Laplace pressure. A proper long-wave form is obtained by approximating the metric factors in the energy functional~\eqref{eq:brushevap-free-energy-functional} as \(\xiz\approx 1+\frac{1}{2}|\nabla\zeta|^2\) and \(\xihz\approx 1+\frac{1}{2}|\nabla(h+\zeta)|^2\). The resulting variations correspond to \cref{eq:brushevap-pressures} with $\xihz=\xiz\approx1$. For a discussion of the merits of the full-curvature form see section~3 of Ref.~\citenum{Thie2018csa}.

Furthermore, the presented three-field model contains several known models as limiting cases:
\begin{enumerate}
  \item \label{en:2field-vol} A  two-field model of a sessile drop of volatile liquid in the narrow gap between two parallel rigid solid plates is obtained in the limit $\zeta\to0$ (and $H_\mathrm{dry}=0$). Technically, one eliminates the second column and second row from the mobility matrices~\eqref{eq:brushevap-conserved-mobility} and~\eqref{eq:brushevap-nonconserved-mobility}, removes $f_\mathrm{brush}$ and all $\zeta$-dependencies from~\eqref{eq:brushevap-free-energy-functional},  thereby reducing the model~\eqref{eq:basics:multi_gradient_dynamics} to two dynamic equations, one for the liquid and one for the vapor. The resulting two-field model is developed and analyzed in Ref.~\citenum{HDJT2023jfm}. It allows for the study of drop evaporation in the full range of parameters from the phase transition-limited case (often covered by ``one-sided models'') to the diffusion-limited case.
\item \label{en:2field-brush} A two-field model of a sessile drop of nonvolatile liquid on an adaptive brush-covered substrate with brush-state dependent interface and wetting energies is obtained in the limit $\rho_\mathrm{vap}\to0$ in the energy functional and the mobilities. Technically, one eliminates the third column and third row from the mobility matrices~\eqref{eq:brushevap-conserved-mobility} and~\eqref{eq:brushevap-nonconserved-mobility}, removes $f_\mathrm{liq}$, $f_\mathrm{vap}$ and $f_\mathrm{air}$ from~\eqref{eq:brushevap-free-energy-functional},  thereby eliminating the dynamic equation for the vapor phase. The resulting model is developed and analyzed in Ref.~\citenum{GrHT2023sm}. Here, we will employ it to study the forced spreading of a drop on a brush-covered substrate.
  \item \label{en:2field-brush-simp} A further simplification -- assuming brush state-independent interface and wetting energies -- results in the more restricted two-field model of Ref.~\citenum{ThHa2020epjt}.
  \item \label{en:1field} Obviously, eliminating columns and rows two and three in the mobility matrices as well as all terms related to vapor and brush in the energy, one obtains (the  full-curvature version of) the basic thin-film model in \cref{eq:onefield:gov}.
  \item \label{en:3field-brush-unif} Assuming a laterally uniform system eliminates all terms with spatial derivatives. The resulting system of ordinary differential equation may be employed to study spatially homogeneous phase change and imbibition dynamics as well as the corresponding steady states, e.g., the sorption isotherm of a brush.
\end{enumerate}

Beside the discussed simplifications, the gradient dynamics form also lends itself conveniently to several extensions of the model: Combining the model with gradient dynamics models for nonvolatile drops covered by insoluble~\cite{ThAP2012pf} or soluble surfactants~\cite{ThAP2016prf} one may study situations where the surfactants do or do not transfer into the brush or/and the gas phase. One could also combine the approach with the one of Ref.~\citenum{ThTL2013prl} and consider drops of mixtures or solutions on brush-covered substrates.  Although such extensions seem complicated because they involve more fields, the overall structure remains very transparent.

\subsection{Equilibrium contact angles}
\label{sec:angles}

\begin{figure}[!tbp]
  \centering
  \includegraphics[width=.6\textwidth]{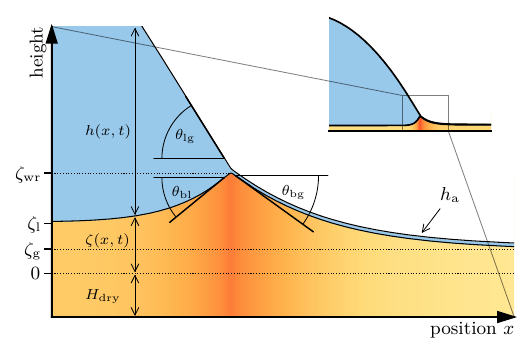}
  \caption{Shown is a sketch of the mesoscopic contact line geometry for a partially wetting liquid on a polymer brush. The indicated equilibrium (Neumann) angles $\theta_\mathrm{lg}$, $\theta_\mathrm{bl}$ and $\theta_\mathrm{bg}$ are measured \bfuwe{with respect to} the horizontal and defined at the respective position of the steepest slope. $\zeta_\mathrm{l}$ and $\zeta_\mathrm{g}$ indicate the equilibrium swelling heights underneath the liquid and underneath the gas phase far away from the contact line region. }\label{fig:contact_angles}
\end{figure}
Finally, we briefly review that in the case of a partially wetting liquid, the free energy Eq.~\eqref{eq:brushevap-free-energy-functional} implies the validity of an (amended) global Young's law and a local Neumann's law at the contact line. In the limiting case of a droplet on a rigid substrate they reduce to the usual Young's law.

The derivation of relations between the asymptotic angles formed by the various interfaces solely relies on the notion of thermodynamic equilibrium, i.e.\ the energy in Eq.~\eqref{eq:brushevap-free-energy-functional} is minimized under the constraint of an imposed total particle number. Hence, all equilibria minimize the grand potential
\begin{equation}
  \mathcal{G}=\mathcal{F}-\int_\Omega \mu (\psi_1 +\psi_2+ \psi_3)\, \mathrm{d}^2r,\label{eq:brushevap-grand-potential}
\end{equation}
where we introduce the chemical potential $\mu$ as a Lagrange multiplier that ensures particle conservation. Therefore, equilibria fulfill the coupled Euler-Lagrange equations $\frac{\delta \mathcal{F}}{\delta \psi_1}=\frac{\delta \mathcal{F}}{\delta \psi_2}=\frac{\delta \mathcal{F}}{\delta \psi_3}=\mu$, where the functional variations are given by Eq.~\eqref{eq:brushevap-pressures}. The third equation is decoupled and easily solved, i.e.\ the system adopts a spatially homogeneous equilibrium vapor concentration that only depends on the chemical potential $\mu$. The remaining system of two coupled equations is more intuitively understood in the hydrodynamic formulation: Because of the constant liquid density we can introduce the pressure $P=\mu \rho_{\mathrm{liq}}$ and obtain $\frac{\delta \mathcal{F}}{\delta h}=\frac{\delta \mathcal{F}}{\delta \zeta}=P$. Therefore, at equilibrium  the polymer brush and the liquid phase share a common spatially homogeneous pressure. This system of two coupled Euler-Lagrange equations is studied in Ref.~\citenum{GrHT2023sm}. There it is shown, that a wetting ridge as sketched in  \cref{fig:contact_angles} forms in the three-phase contact line region. It can be clearly identified if the brush swells on an intermediate scale, i.e.\ if the swelling is large compared to the microscopic scale formed by the adsorption layer \bfuwe{thickness} and small compared to the drop size. Then, the global equilibrium contact angle $\theta_Y=\theta_\mathrm{lg}$ is governed by an amended Young's law, namely,
\begin{equation}
  \gamma_{\mathrm{lg}}\cos\theta_\mathrm{Y} = \gamma_\mathrm{lg}+f_\mathrm{wet}(h_\mathrm{a}, \zeta_\mathrm{g})
  + \gamma_\mathrm{bl}(\zeta_\mathrm{g}) + g_\mathrm{brush}(\zeta_\mathrm{g}) -  \gamma_\mathrm{bl}(\zeta_\mathrm{l}) - g_\mathrm{brush}(\zeta_\mathrm{l}).\label{eq:young}
\end{equation}
As the brush swelling $\zeta_l$ and $\zeta_g$ underneath the liquid film and underneath the vapor phase, respectively, may have different equilibrium values, the law contains the two additional brush energy-dependent contributions $g_\mathrm{brush}(\zeta_\mathrm{l})$ and $g_\mathrm{brush}(\zeta_\mathrm{g})$. Therefore, the global contact angle $\theta_\mathrm{Y}$, measured on a scale where wetting ridge-related details of the contact line region are not visible, differs from the case of a drop on a rigid \bfuwe{inert} substrate. Although, the mesoscopic relation~\eqref{eq:young} contains the wetting energy $f_\mathrm{wet}(h_\mathrm{a}, \zeta_\mathrm{g})$ it is fully consistent with the analogous macroscopic global Young's law if one \bfuwe{identifies} the mesoscopic expression $\gamma_\mathrm{lg}+f_\mathrm{wet}(h_\mathrm{a}, \zeta_\mathrm{g}) + \gamma_\mathrm{bl}(\zeta_\mathrm{g})$ \bfuwe{as representing} the macroscopic brush-gas interface energy $\gamma_\mathrm{bg}(\zeta)$.

In the contact line region where the wetting ridge is situated a local Neumann law holds. Its derivation is presented in section~3.2.3 of Ref.~\citenum{GrHT2023sm}. Denoting the brush height at the tip of the wetting ridge by $H_\mathrm{dry}+\zeta_\mathrm{wr}$ and defining the appropriate angles as indicated in \cref{fig:contact_angles} it reads
\begin{equation}
  \begin{split}
    \gamma_\mathrm{lg}\cos\theta_\mathrm{lg}+\gamma_\mathrm{bl}(\zeta_\mathrm{wr})\cos{\theta_\mathrm{bl}}=[\gamma_\mathrm{lg}+\gamma_\mathrm{bl}(\zeta_\mathrm{wr})+f_\mathrm{wet}(h_\mathrm{a},\zeta_\mathrm{wr})]\cos\theta_\mathrm{bg}	\\
    \gamma_\mathrm{lg}\sin\theta_\mathrm{lg}+\gamma_\mathrm{bl}(\zeta_\mathrm{wr})\sin\theta_\mathrm{bl}=[\gamma_\mathrm{lg}+\gamma_\mathrm{bl}(\zeta_\mathrm{wr})+f_\mathrm{wet}(h_\mathrm{a}, \zeta_\mathrm{wr})]\sin\theta_\mathrm{bg}, \label{eq:Neumann}
  \end{split}
\end{equation}
As expected, the law is rotation invariant, i.e.\ it  only determines the contact angles relative to each other but not with respect to the horizontal. For the present case of a brush, this invariance is broken by the identification $\theta_\mathrm{lg}=\theta_Y$, i.e.\ the angle $\theta_\mathrm{lg}$ is already defined by the global Young's law~\eqref{eq:young}. The two remaining equilibrium angles $\theta_\mathrm{bl}$ and $\theta_\mathrm{bg}$ are then fully determined by Eqs.~\eqref{eq:Neumann}. A macroscopic version of Neumann's law could be written with a similar consistency condition as above, but would contain a wetting ridge height-dependent liquid-gas interface energy. Below we will use Neumann's law to distinguish different dynamic regimes of contact lines moving over brush-covered substrates.

\section{Selected examples}
\label{sec:examples}

In the following we consider six examples for the usage of the developed model(s). We start with the discussion of sorption isotherms.

\subsection{Sorption isotherm of a polymer brush}\label{sec:brushevap-isotherm}
First, we consider the limiting case~\ref{en:3field-brush-unif}, namely, the sorption isotherm of a brush obtained when considering the general model with eliminated spatial derivatives. Then, the variations~\eqref{eq:brushevap-pressures} (chemical potentials) are
\begin{equation}
  \begin{aligned}
    \frac{\delta \mathcal{F}}{\delta \psi_1}  & = \frac{1}{\rho_\mathrm{liq}} \left[f_\mathrm{liq} + \partial_h f_\mathrm{wet}(h, \zeta)\right]\\
    \frac{\delta \mathcal{F}}{\delta \psi_2} & = \frac{1}{\rho_\mathrm{liq}} \bigg\{ \partial_\zeta \left[ \gamma_\mathrm{bl}(\zeta) + \f{wet}(h,\,\zeta) + \f{brush}(\zeta) \right] + \f{liq} \bigg\}\\
    \frac{\delta \mathcal{F}}{\delta \psi_3}   & = k_B T \log \left( \frac{\rho_\mathrm{vap}}{\rho_\mathrm{tot} - \rho_\mathrm{vap}} \right).
  \end{aligned}
\end{equation}
At equilibrium all chemical potentials balance and the transfer fluxes \(j_\mathrm{ev}\), \(j_\mathrm{ev}'\), and \(j_\mathrm{im}\) are zero. We may consider either brush-liquid coexistence, i.e.\ the liquid film is very thick ($f_\mathrm{wet}\to0$ and $\gamma_\mathrm{bl}= \gamma_\mathrm{bl}(\zeta_\mathrm{l})$) or brush-vapor coexistence, i.e.\ the liquid film corresponds to the adsorption layer of \bfuwe{thickness} $\ha$, und $f_\mathrm{wet}\to f_\mathrm{wet}(\ha, \zeta_\mathrm{g})$ giving an effective brush-vapor interface energy $\gamma_\mathrm{bg}(\zeta_\mathrm{g}) = \gamma_\mathrm{bl}(\zeta_\mathrm{g}) + \f{wet}(\ha,\zeta_\mathrm{g}) + \gamma_\mathrm{lg}$.

In the case of brush-liquid coexistence, a thick uniform liquid layer is at the same time in equilibrium with an ambient vapor phase and the brush, i.e.\ independently of the brush state the air is saturated with vapor. Balancing the chemical potentials in liquid layer and vapor, \(\delta \mathcal{F}/\delta \psi_1 = \delta \mathcal{F}/\delta \psi_3\), gives
\begin{align}
  \frac{\f{liq}}{\rho_\mathrm{liq}} & = k_B T \log \left( \frac{\rho_\mathrm{sat}}{\rho_\mathrm{tot} - \rho_\mathrm{sat}} \right).\label{eq:brushevap-sat_cond}
\end{align}

Similarly, the thick liquid layer coexists with the homogeneously swollen brush. Balancing the chemical potentials of film and brush, \(\delta \mathcal{F}/\delta \psi_1 = \delta \mathcal{F}/\delta \psi_2\), gives
\begin{equation}
  \partial_\zeta \left[ \gamma_\mathrm{bl}(\zeta_\mathrm{l}) + \f{brush}(\zeta_\mathrm{l}) \right] = 0.\label{eq:eq_liquid_bulk}
\end{equation}
In other words, the minimization of the combined brush bulk energy and brush-liquid interface energy defines the equilibrium thickness of the brush. Note that the condition also holds in the case of a very large drop as then the Laplace pressure approaches zero.

In contrast, in the case of brush-vapor coexistence, the resulting condition is
\begin{equation}
  \partial_\zeta \left[ \gamma_\mathrm{bg}(\zeta_\mathrm{g}) + \f{brush}(\zeta_\mathrm{g}) \right] + \f{liq} = k_B T \rho_\mathrm{liq} \log \left( \frac{\rho_\mathrm{vap}}{\rho_\mathrm{tot} - \rho_\mathrm{vap}} \right)
  \label{eq:brush-vap-cond}
\end{equation}
where we used that $\gamma_\mathrm{lg}$ does not depend on $\zeta$. This gives the brush state in coexistence with the vapor \(\zeta_\mathrm{g}\) as a function of vapor density \(\rho_\mathrm{vap}\). Using the saturation condition~\eqref{eq:brushevap-sat_cond} to replace $\f{liq}$ in~\eqref{eq:brush-vap-cond} we obtain
\begin{align}
  \partial_\zeta \left[ \gamma_\mathrm{bg}(\zeta_\mathrm{g}) + \f{brush}(\zeta_\mathrm{g}) \right]
  & =\rho_\mathrm{liq} k_B T \log \left( \frac{\rho_\mathrm{vap}}{\rho_\mathrm{tot} - \rho_\mathrm{vap}}  \frac{\rho_\mathrm{tot} - \rho_\mathrm{sat}}{\rho_\mathrm{sat}} \right) \nonumber\\
  & \approx\rho_\mathrm{liq} k_B T \log \left( \frac{\rho_\mathrm{vap}}{\rho_\mathrm{sat}} \right),
  \label{eq:brush-sorption-full}
\end{align}
where in the last step we have used, that the total particle density of the gas is much larger than the vapor particle density, \(\rho_\mathrm{vap}<\rho_\mathrm{sat}\ll\rho_\mathrm{tot}\). The argument of the logarithm is a relative saturation (or relative humidity) \(\phi=\rho_\mathrm{vap}/\rho_\mathrm{sat}\). Such a relation between the relative vapor concentration and the liquid uptake of a medium, here the polymer brush, is called the \emph{sorption isotherm}~\cite{RiCD2022aapm}.

\begin{figure}[tb]
  \centering
  \includegraphics{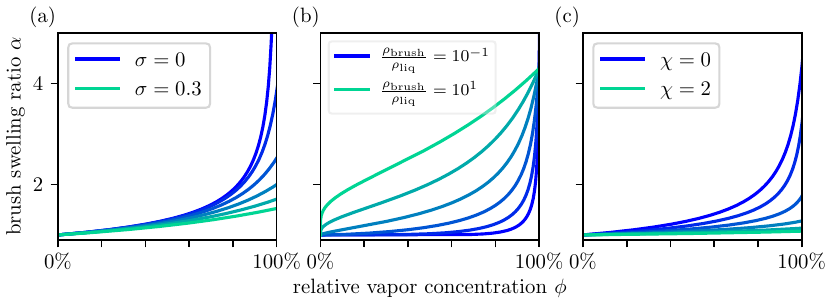}
  \caption{Sorption isotherms $\alpha(\phi)$ for a polymer brush in contact with vapor according to \cref{eq:brushevap-iso2} for various parameter configurations. Panels (a), (b) and (c) show how the isotherm changes with increasing grafting density $\sigma$, the density ratio \(\rho_\mathrm{brush}/\rho_\mathrm{liq}\), and the Flory parameter \(\chi\), respectively. The respective other parameters are fixed at \(\sigma=\num{0.05}\), \(\rho_\mathrm{brush}/\rho_\mathrm{liq}=1\), and \(\chi=0\). In (a) and (c) equidistant steps in the varying parameter are used while (b) uses logarithmic steps.}\label{fig:brushevap-isotherm}
\end{figure}

If one assumes that the brush is very thick (i.e.\ its energy per substrate area is much larger than the interface energy), one may neglect the interface contributions in~\eqref{eq:brush-sorption-full}. Then, expressing the sorption isotherm in terms of the relative humidity $\phi$ and the brush swelling ratio $\alpha=(H_\mathrm{dry} + \zeta_\mathrm{g})/H_\mathrm{dry}$, one has
\begin{equation}
  \phi = \exp \left\{ \frac{\rho_\mathrm{brush}}{\rho_\mathrm{liq}} \left[ \sigma^2 \alpha + 1/\alpha + \log (1-1/\alpha) + \chi /\alpha^2 \right] \right\}
  \label{eq:brushevap-iso2}
\end{equation}
in agreement with the literature~\cite{RVNB2020m, RiCD2022aapm}. \bfuwe{Note that the first term in the exponential on the r.h.s.\ of Eq.~\eqref{eq:brushevap-iso2} arises from the elastic contribution in the Alexander-de Gennes mean-field approach. It is absent if a non-grafted polymer layer is considered~\cite{LBHS1999m, DVRL2013l}. In consequence of the final approximation the sorption isotherm is solely related to $\partial_\zeta f_\mathrm{brush}$. Therefore, further refinements of the brush bulk energy $f_\mathrm{brush}$ allow for adjustments of the sorption isotherm and vice versa. As a further consequence of this approximation, the swelling of the brush in fully vapor-saturated gas (\(\phi=1\)) is identical to the swelling of a brush in contact with the liquid. Examples of sorption isotherms according to \cref{eq:brushevap-iso2} for various parameter configurations are shown in \cref{fig:brushevap-isotherm}.}
However, if the brush state-dependent interface energies $\gamma_\mathrm{bl}(\zeta)$ and $\gamma_\mathrm{bg}(\zeta)$ are different and of a similar order of magnitude as the brush energy $\f{brush}(\zeta)$, the above approximation can not be used. Then, the state of a brush coexisting with bulk liquid necessarily differs from the state of a brush coexisting with saturated vapor \bfuwe{(see discussion below).}

\begin{figure}[tb]
  \centering
  \includegraphics[width=0.8\hsize]{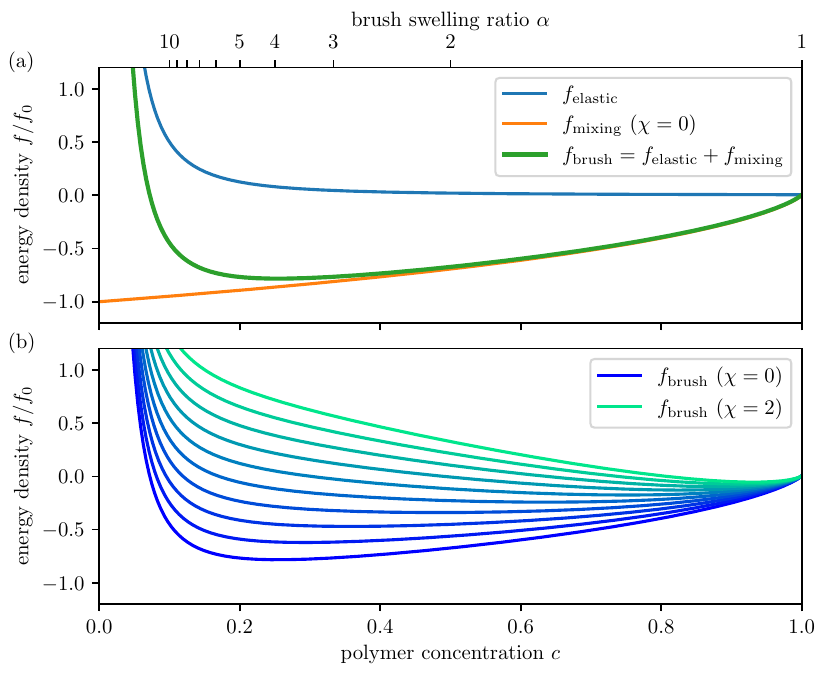}
  \caption{Panel (a) illustrates for $\chi=0$ the interplay of elastic component $f_\mathrm{elastic}$ and mixing component $f_\mathrm{mixing}$ within the brush energy $\f{brush}$ (\cref{eq:brush-f_brush}). Shown is the dependence on the polymer concentration \(c=H_\mathrm{dry}/(H_\mathrm{dry}+\zeta)\) (lower $x$-axis) and swelling ratio $\alpha=1/c$ (upper $x$-axis). Panel~(b) shows how the Flory miscibility parameter \(\chi\) influences the brush energy \(\f{brush}\). The employed grafting density is \(\sigma=\num{0.1}\). All energies are given as density per substrate area, and in units of \(f_0 = H_\mathrm{dry}\rho_\mathrm{brush} k_B T\).}\label{fig:brush-energy}
\end{figure}

To better understand the interplay of elastic component $f_\mathrm{elastic}$ and mixing component $f_\mathrm{mixing}$ within the brush energy $\f{brush}$, in \cref{fig:brush-energy}~(a) we show how for $\chi=0$ the single minimum of  $\f{brush}(c)$ results from the interplay of a monotonically decreasing $f_\mathrm{elastic}$ and monotonically increasing $f_\mathrm{mixing}$. In other words, the mixing contribution favors swollen brushes while the elastic \bfuwe{contribution penalizes} strong swelling. The grafting density \(\sigma\) is fixed at \(\sigma=\num{0.1}\) and scales the elastic energy, which slightly shifts the location of the minimum.

The mixing contribution strongly depends on the Flory parameter \(\chi\). \Cref{fig:brush-energy}~(b) illustrates how the value of $\chi$ influences the dependence of the brush energy \(\f{brush}\) on polymer concentration \(c\) (lower $x$-axis) and swelling ratio $\alpha$ (upper $x$-axis). For a good solvent (\(\chi=0\)) one finds the pronounced minimum of \cref{fig:brush-energy}~(a). If, however, the miscibility of polymers and liquid decreases (\(\chi>0\)) swelling of the brush is energetically less favored, i.e.\ the minimum of \(\f{brush}\) shifts to higher polymer concentrations and becomes more shallow, see \cref{fig:brush-energy}~(b).

\begin{figure}[tb]
  \centering
  \includegraphics[width=0.8\hsize]{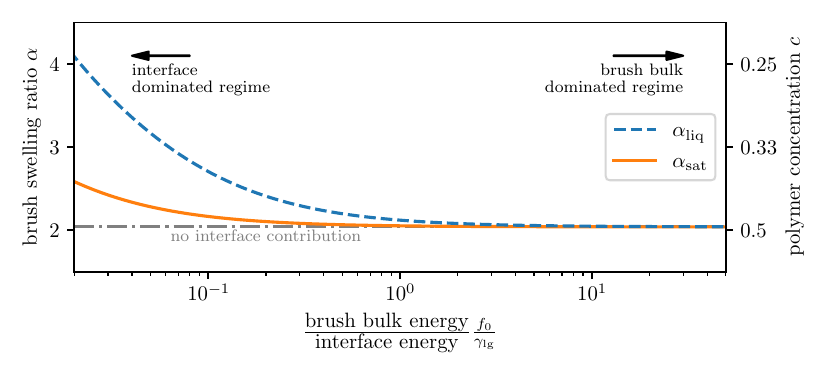}
  \caption{Shown is the equilibrium swelling of a brush in contact with liquid $\alpha_\mathrm{liq}$ in comparison to the swelling of a brush in contact with saturated vapor $\alpha_\mathrm{sat}$ as a function of the ratio $f_0/\gamma_\mathrm{lg}$ of the brush bulk energy density $f_0 = H_\mathrm{dry}\rho_\mathrm{brush} k_B T$ and the liquid-gas interface energy density $\gamma_\mathrm{lg}$. The remaining parameters are $\sigma=0.3$, $\chi=0$, $\gamma_\mathrm{lg}=27 \mathrm{mN/m}$, $\gamma_\mathrm{bl,dry}=3 \mathrm{mN/m}$ and the (dry brush) Young angle is set to $\thetaeq=\SI{25}{\degree}$. }
  \label{fig:sorption_liq_vs_vapor}
\end{figure}

Finally, we stress \bfuwe{the mentioned} important consequence of the dependence of the interface energies on brush state that we introduced in Eqs.~\eqref{eq:brush-f_wet_new} and \eqref{eq:brush-gamma-deps}: A comparison of the equilibrium swelling states of a brush in direct contact with a bulk liquid phase $\alpha_\mathrm{liq}$ and of the same brush in contact with a vapor phase at saturation $\alpha_\mathrm{sat}$ shows that these swelling states are not necessarily identical. In the case of coexistence with saturated vapor the r.h.s.\ of Eq.~\eqref{eq:brush-sorption-full} vanishes, i.e.\ one has
\begin{equation}
  \partial_\zeta \left[ \gamma_\mathrm{bg}(\zeta) + \f{brush}(\zeta) \right]=0,
  \label{eq:brush-saturated}
\end{equation}
i.e.\ the equilibrium brush state in coexistence with saturated vapor, $\zeta_\mathrm{sat}$, is given by the minimum of the sum of brush energy and brush-gas interface energy.

\Cref{eq:brush-saturated} is of similar form as \cref{eq:eq_liquid_bulk}, that gives the equilibrium brush state in coexistence with bulk liquid  $\zeta_\mathrm{l}$, as the minimum of the sum of brush energy and brush-liquid interface energy. This implies that $\zeta_\mathrm{sat}$ only equals $\zeta_\mathrm{l}$ if one can either neglect the interface energies as compared to the brush energy or if the interface energies $\gamma_\mathrm{bg}$ and $\gamma_\mathrm{bl}$ have the same functional dependence on $\zeta$. The latter condition ($\partial_\zeta \gamma_\mathrm{bl}=\partial_\zeta \gamma_\mathrm{bg}$) does not even hold for our simplistic linear interpolation between the limiting cases (i.e.\ setting all exponents in \cref{eq:brush-gamma-deps} equal to one). Hence, in general, the brush equilibrium swellings underneath the liquid and underneath the saturated vapor are noticeably different, if the magnitude of the interface energy is comparable to the one of the brush bulk energy.

This is illustrated in \cref{fig:sorption_liq_vs_vapor}, where we give the swelling ratios $\alpha_\mathrm{liq}$ and $\alpha_\mathrm{vap}$ as a function of the ratio $f_0/\gamma_\mathrm{lg}$ between the typical scales of brush energy $f_0=H_\mathrm{dry}\rho_\mathrm{brush} k_B T$ and interface energy. We find that in the brush bulk-dominated case, i.e.\ if the ratio is large, nearly no difference between $\alpha_\mathrm{liq}$ and $\alpha_\mathrm{vap}$ is discernible. In contrast, if the interface energy dominates, i.e.\ if \bfuwe{$f_0/\gamma_\mathrm{lg}$} is small, $\alpha_\mathrm{liq}$ and $\alpha_\mathrm{vap}$ can strongly differ. For instance, at a ratio of $f_0/\gamma_\mathrm{lg}=0.1$, \cref{fig:sorption_liq_vs_vapor} indicates that $\alpha_\mathrm{liq}$ is approximately 30\% larger than $\alpha_\mathrm{sat}$. It is to be expected that the effect is more pronounced if the dependencies of $\gamma_\mathrm{bl}$ and $\gamma_\mathrm{bg}$ on $\zeta$ are nonlinear.

Such a difference might at first seem counter-intuitive, as normally one only considers bulk phases when discussing thermodynamic coexistence. However, here, one of the coexisting phases -- the thin brush layer -- is itself of reduced dimension. This implies that brush and interface energies might be of a similar scale resulting in a notable difference between brush swelling under bulk liquid and under its saturated vapor. A seemingly similar effect is experimentally observed~\cite{RiCD2022aapm,OSCB2015p} and sometimes referred to as the Schr\"oder effect~\cite{VWRK2006jms,DavP2014zpcjrpccp}. However, it remains a question for the future to quantitatively compare the effect discussed here with specific experimental measurements. \bfuwe{In particular, it would be interesting to see whether the implied brush-thickness dependence of a Schr\"oder effect for polymer brushes is found in experiments.}

\subsection{Relaxational spreading dynamics}\label{sec:brush-spreading}

Next we employ the limiting case~\ref{en:2field-brush}, i.e.\ the dynamical model~\eqref{eq:brushcons-fluxes} without the vapor part, to study the spreading of a droplet on an initially dry polymer brush by numerical simulation.\footnote{In particular, we use the finite element method implemented in the package \texttt{oomph-lib}~\cite{HeHa2006}.} The simulation is initialized with a drop state that corresponds to the equilibrium for a nonadaptive substrate with a Hamaker constant \(A_\mathrm{dry}\). The polymer brush is slightly pre-swollen (nearly dry, \(\alpha(t=0)=1/c(t=0)=\num{1.1}\)), as the brush pressure $\delta \mathcal{F}/\delta \psi_2$ diverges for a completely dry brush. Then, the simulation starts when at \(t=0\) the liquid-brush interaction is ``switched on'', i.e.\ the substrate is allowed to absorb liquid from the drop. The simulation is performed using spatial and temporal adaptivity to resolve both the wetting ridge and the relaxational dynamics. Consequently, when the adaptive time step diverges to infinity, the system has reached its equilibrium state.

We assume a radially symmetric geometry, transform the model into polar coordinates, employ corresponding boundary conditions, and perform the simulations on the resulting one-dimensional radial domain \(\mathcal{D}_r=[0,\,L]\) (for details see section 3.1.2 of Ref.~\citenum{Hartmann2023Munster}). We employ the full-curvature formulation.\footnote{Although we use the rather small ratio of adsorption layer thickness and drop height of \(\ha / h(r=0,\,t=0) = 1/1000\), the dynamics can exhibit a significant (and unintended) flow of liquid through the adsorption layer, bypassing the typically slower diffusive transport within the brush layer. In consequence, the brush swells unphysically fast even in areas far away from the drop. We prevent this effect by modulating the (nonconserved) transfer rate \(M\) of liquid exchange between the brush and the adsorption layer employing a smooth step function with threshold and tolerance values of \(h_\mathrm{thresh}=1.1\,\ha\) and \(h_\mathrm{tol}=\num{0.04}\,\ha\), respectively.
}

\begin{figure}[!tbp]
  \centering
  \includegraphics[width=\textwidth]{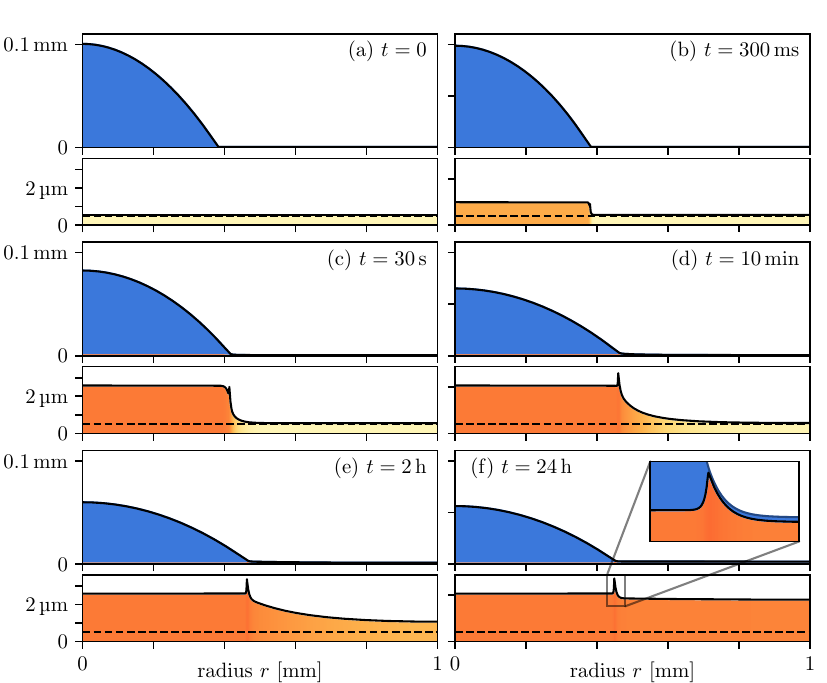}
  \caption{Snapshots from a typical spreading dynamics  of a drop of partially wetting nonvolatile liquid on an initially dry polymer brush at selected times as given in the individual panels. The height profiles of the liquid drop \(h+H\) and of the brush \(H=H_\mathrm{dry}+\zeta\) are shown as solid lines while the liquid concentration \(1-c=1-H_\mathrm{dry}/H\) in the brush is indicated by a yellow to orange shading of the brush layer. The respective upper [lower] panels show both, the drop and (comparably thin) brush layer [a magnification of the brush layer]. The horizontal dashed line indicates the reference height \(H_\mathrm{dry}\) of a completely dry brush.  Panel (f) corresponds to the final equilibrium state, i.e.\ the limit \(t\to\infty\). Corresponding dependencies of selected quantities on time are presented in \cref{fig:brush-spreading-contactangles}. The employed parameters represent water at normal conditions (see section 3 of Ref.~\citenum{Hartmann2023Munster}), \(\thetaeq=\SI{30}{\degree}\), \(\ha=\SI{0.1}{\micro m}\) (\(\approx 1/1000\) of drop height), and an exemplary brush with \(\sigma=0.1\), \(H_\mathrm{dry}=\SI{500}{nm}\), \(\ell_K = \SI{4}{nm}\), \(\gamma_\mathrm{bl} = \SI{20}{mN/m}\), \(D_\text{brush}=\SI{e-7}{m^2/s}\) and \(M_\text{im}=\SI{e-10}{m/Pa\,s}\). An enlarged visualization of the wetting ridge is shown in the inset of panel (f).
  }\label{fig:brush-spreading-snapshots}
\end{figure}

The typical spreading dynamics is illustrated in \cref{fig:brush-spreading-snapshots} by snapshots of drop and brush state at six selected times (stated in the panels). As the height of the polymer brush is quite small compared to the height of the drop, it is barely visible in the respective upper panels. Therefore, the respective \bfuwe{lower} panels give the brush profile on a strongly magnified height scale. Additionally, the absorbed amount of liquid within the brush \(1-c=1+\zeta/H_\mathrm{dry}\) is encoded by the yellow (\(c=1\)) to orange (\(c\ll 1\)) \bfuwe{shading.} A color bar is not provided, as the full information on the brush state is already captured by its height profile.

\begin{figure}[!tbp]
  \centering
  \includegraphics{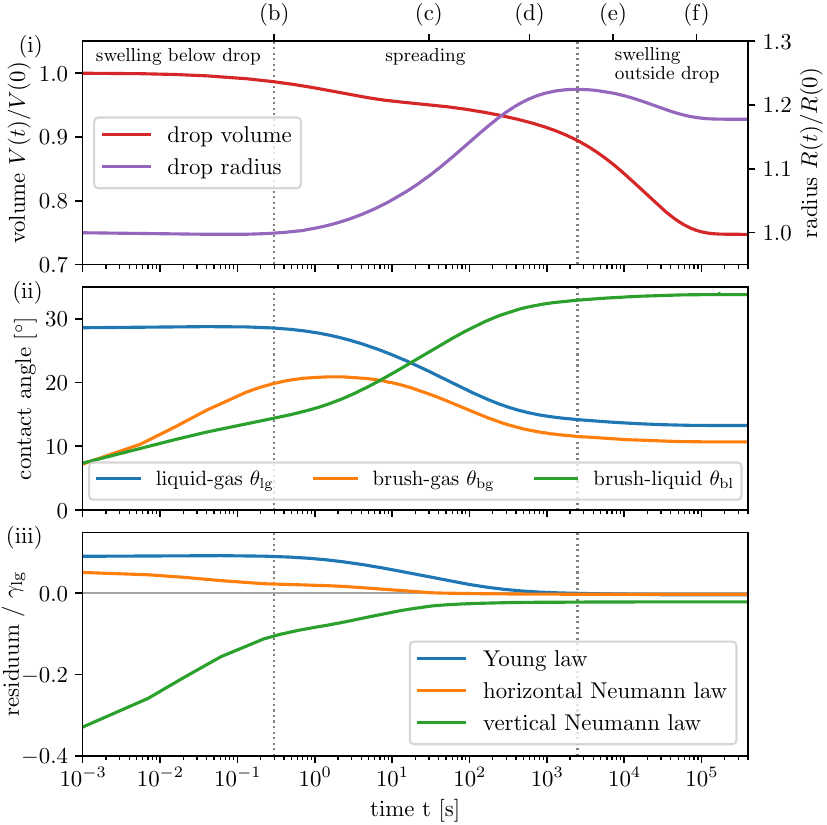}
  \caption{
    For the spreading drop illustrated in \cref{fig:brush-spreading-snapshots} time dependencies are presented of (i)~the drop volume \(V(t)\) and radius \(R(t)\), (ii)~the measured Neumann contact angles \(\theta_\mathrm{lg}\),  \(\theta_\mathrm{bl}\), \(\theta_\mathrm{bg}\), and (iii) the residuals of the mesoscopic Young and Neumann laws obtained with the measured angles, to quantify how well the laws are fulfilled.
    The borders between regimes of dominant swelling below the drop, dominant drop spreading, and dominant swelling outside the drop are indicated by vertical dotted lines. The times corresponding to the snapshots in \cref{fig:brush-spreading-snapshots}~(b)--(f) are indicated at the top of panel~(i). Note that the time axis is logarithmic.
  }\label{fig:brush-spreading-contactangles}
\end{figure}

Beside the visual impression in \cref{fig:brush-spreading-snapshots} we provide in \cref{fig:brush-spreading-contactangles} the corresponding time evolutions of several quantities: Panel~(i) gives the drop volume and radius, panel~(ii) gives the three Neumann angles \(\theta_\mathrm{lg}\),  \(\theta_\mathrm{bl}\), and \(\theta_\mathrm{bg}\) obtained from the steepest slopes of liquid-gas, brush-liquid and brush-gas interfaces in the region  of three phase contact, and panel~(iii) quantifies how well the \bfuwe{global} Young law~\eqref{eq:young} and the Neumann law~\eqref{eq:Neumann} are fulfilled during each stage of the dynamics.\footnote{The residuals of the mesoscopic Young and Neumann laws according to Eqs.~\eqref{eq:young} and~\eqref{eq:Neumann} are defined as
  \begin{align}
    \text{res}_\text{Young} = &\ \gamma_\mathrm{lg}+\f{wet}(\ha,\,\zeta_\mathrm{g}) - \gamma_\mathrm{lg}\cos\theta_\mathrm{lg}
    + \gamma_\mathrm{bl}(\zeta_\mathrm{g}) + \f{brush}(\zeta_\mathrm{g}) -  \gamma_\mathrm{bl}(\zeta_\mathrm{l}) - \f{brush}(\zeta_\mathrm{l}),\nonumber\\
    \text{res}_\text{Neumann,horizontal}=&[\gamma_\mathrm{lg}+\gamma_\mathrm{bl}(\zeta_\mathrm{wr})+\f{wet}(\ha,\,\zeta_\mathrm{wr})]\cos\theta_\mathrm{bg}
    -\gamma_\mathrm{lg}\cos\theta_\mathrm{lg}-\gamma_\mathrm{bl}(\zeta_\mathrm{wr})\cos{\theta_\mathrm{bl}}, \nonumber\\
    \text{res}_\text{Neumann,vertical}=&[\gamma_\mathrm{lg}+\gamma_\mathrm{bl}(\zeta_\mathrm{wr})+\f{wet}(\ha,\,\zeta_\mathrm{wr})]\sin\theta_\mathrm{bg}
    -\gamma_\mathrm{lg}\sin\theta_\mathrm{lg}-\gamma_\mathrm{bl}(\zeta_\mathrm{wr})\sin\theta_\mathrm{bl}. \nonumber
  \end{align}
  where the angles and brush states represent the actual time-dependent quantities measured during the time evolution. The angles in \cref{fig:brush-spreading-contactangles}~(ii) are given in degrees as we employ the full-curvature form of the model.}
The time scale is logarithmic allowing us to highlight three qualitatively different phases of the spreading dynamics where different transport processes dominate as discussed next.

The initial state pictured in panel~(a) of \cref{fig:brush-spreading-snapshots} starts to relax after we ``switch on'' the substrate interaction: Within the first few milliseconds we observe a rapid swelling of the brush directly underneath the drop, see panel (b). The droplet volume $V$ remains mostly constant, as only a small amount of liquid imbibes the brush. Drop radius $R$ and liquid-gas contact angle $\theta_\mathrm{lg}$ also remain constant, cf.~\cref{fig:brush-spreading-contactangles}. With advancing time, the swelling below the drop continues while other slower dynamical processes start to contribute. Most notably, within the first few seconds (cf.~\cref{fig:brush-spreading-contactangles}~(c)), a wetting ridge forms in the three phase contact region, i.e.\ the brush-liquid and brush-gas contact angles increase ($\theta_\mathrm{bl}$ and $\theta_\mathrm{bg}$), see \cref{fig:brush-spreading-contactangles}~(ii). In parallel, $R$ starts to grow and $\theta_\mathrm{lg}$ decreases as the drop spreads. At the same time the imbibition dynamics induces brush swelling also in the region adjacent to the drop, thereby absorbing more and more liquid. Due to this swelling, $\theta_\mathrm{bg}$ starts to decrease again as the formation of a swelling ``halo'' relaxes the initially steep slope of the brush profile.

The spreading and ridge formation continue for some minutes and reach their final form within the first \(10-60\) minutes, corresponding to \cref{fig:brush-spreading-snapshots}~(d). Beyond this time, only subtle changes in the contact angles occur, while the radius actually starts to shrink as due to the radial geometry the brush far away from the drop absorbs a large volume of liquid. This occurs on a time scale of hours (see panel~(e)). After about 1 day the brush is fully swollen and the system has reached its equilibrium state of a steady drop coexisting with a swollen brush, see panel~(f). Note that the wetting ridge persists even when all dynamic processes have come to a halt. We remark that the time scales of imbibition dynamics (absorption and wicking) strongly (and linearly) depend on the values of the transport coefficients (transfer constant \(M\) and the diffusion coefficient \(D\), respectively). In the discussed example their values are chosen such that absorption is much faster than diffusion.

\cref{fig:brush-spreading-contactangles}~(iii) shows the residuals of the mesoscopic Young law~\eqref{eq:young} and Neumann law~\eqref{eq:Neumann} obtained when using them with the measured dynamic angles. In this way we quantify the deviation of the dynamic angles from the equilibrium laws and their approach in the long-time limit. Thereby, the residuals are given in units of  \(\gamma_\mathrm{lg}\). At equilibrium, the Young and horizontal Neumann laws perfectly hold [\(\text{res}_\text{Young}\approx\num{-4e-3}\) and  \(\text{res}_\text{Neumann,horizontal}\approx\num{-2e-3}\)], only the vertical Neumann law shows a small visible deviation [\(\text{res}_\text{Neumann,vertical}\approx\num{-2e-2}\)]. This is a consequence of the smoothening of the macroscopically sharp wetting ridge in the mesoscopic picture.

\begin{figure}[!tbp]
  \centering
  \includegraphics{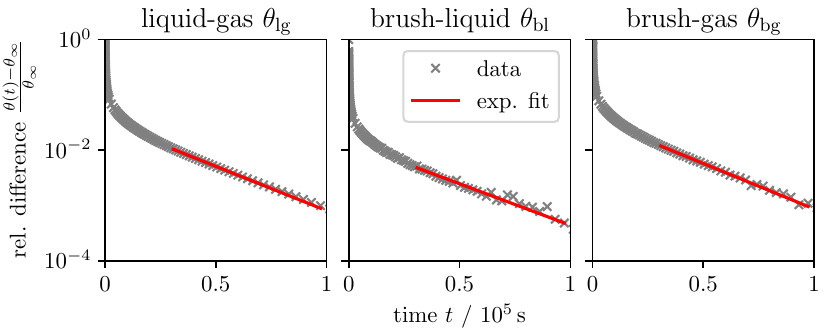}
  \caption{Log-normal plot of three Neumann angles \(\theta_\mathrm{lg}(t)\),  \(\theta_\mathrm{bl}(t)\), and \(\theta_\mathrm{bg}(t)\) as a function of time showing the exponential approach to their equilibrium values \(\theta(t\to\infty)=\theta_\infty\). The measured values (crosses) may be approximated by an exponential fit \(f(t)=\lambda \exp(-t/\tau)\) as indicated by the red solid lines. }\label{fig:brush-exp-decay}
\end{figure}

Reference~\citenum{BBSV2018l} assumes that after placing a drop on an adaptive substrate, the interfacial energies exponentially relax towards a new equilibrium value. This implies that Neumann angles should also show such a relaxation. However, Ref.~\citenum{{BBSV2018l}} also mentions that the observed change may be more complex and will depend on the particular considered system. Our simulations confirm both aspects. On the one hand, we find a rather complex dynamics of the angles at short and intermediate times (see \cref{fig:brush-spreading-contactangles} and discussion above) that reflects the interplay of the coupled transport processes. On the other hand, the final equilibrium is indeed approached via an exponential relaxation as evidenced in the log-normal plot of the time-dependence of the three Neumann angles in \cref{fig:brush-exp-decay}.
The values obtained from the simulation \bfuwe{are well approximated\footnote{At large times, the difference to the equilibrium values becomes becomes very small. This results in a high sensitivity towards numerical grid effects and a certain loss of numerical precision becomes visible.} by an exponential fit in the relaxation range \(3\times{}10^4<t<10^5\).}

\subsection{Forced spreading of drops}\label{sec:brush-inflation}
A common experimental technique to probe substrates covered by polymer brushes, hydrogels or other adaptive and/or flexible layers is the study of sessile droplets that undergo inflation-deflation cycles, i.e.\ liquid is pumped in or out of the drop thereby inducing an advancing or receding movement of the contact line, respectively~\cite{WMYT2010scc,SHNF2021acis}. A typical setup employing a syringe pump is, e.g., sketched in Fig.~4 of Ref.~\citenum{SHNF2021acis}. Other examples where drops are inflated in this way to investigate advancing contact lines on substrates with rigid or soft organic coatings are found in Refs.~\citenum{LWLH2002acis,GCRB2004l,TYYA2006l,KDNR2013sm,PBDJ2017sm}. Sometimes, the setup is referred to as dynamic one-cycle contact angle or dynamic cycling contact angle (DCCA) experiment depending on the number of used inflation-deflation cycles~\cite{LWLH2002acis}.

Here, we investigate a similar inflation-deflation cycle employing again the limiting case~\ref{en:2field-brush}, i.e.\ the dynamical model~\eqref{eq:brushcons-fluxes} without the vapor part as in the previous section, but adding an external forcing. We adapt our model by imposing an in- or out-flux \(q\) employing a flux boundary condition for the liquid layer at the apex of the drop.\footnote{Strictly speaking the flux is imposed at a very small radius $R_a$. The modeling domain is therefore restricted to the annular domain \(\mathcal{D}_r=[R_a,\,R_b]\) with \(0<R_a<R_b\).
Within the brush layer, we consider no flux through the domain boundaries, such that the full set of boundary conditions reads:
\begin{align}
  \frac{h^3}{3\eta} \partial_r \frac{\delta \mathcal{F}}{\delta h}\bigg|_{r=R_a} = -q,\qquad \partial_r \frac{\delta \mathcal{F}}{\delta \zeta}\bigg|_{r=R_a} = \partial_r \frac{\delta \mathcal{F}}{\delta h}\bigg|_{r=R_b} = \partial_r \frac{\delta \mathcal{F}}{\delta \zeta}\bigg|_{r=R_b} = 0,\nonumber
\end{align}
with the homogeneous Neumann conditions
\begin{align}
  \partial_r h \Big|_{r=R_a} =
  \partial_r h \Big|_{r=R_b} =
  \partial_r \zeta \Big|_{r=R_a} =
  \partial_r \zeta \Big|_{r=R_b} = 0. \nonumber
\end{align}
}
The system is of interest as within a cycle phases of forced wetting and forced dewetting occur. Thereby, the brush irreversibly adapts its wetting properties as it fills with liquid in the wetting phase. The employed geometry and procedure closely follows the experimentally used protocol.

\begin{figure}[!tbp]
  \centering
  \includegraphics[width=\textwidth]{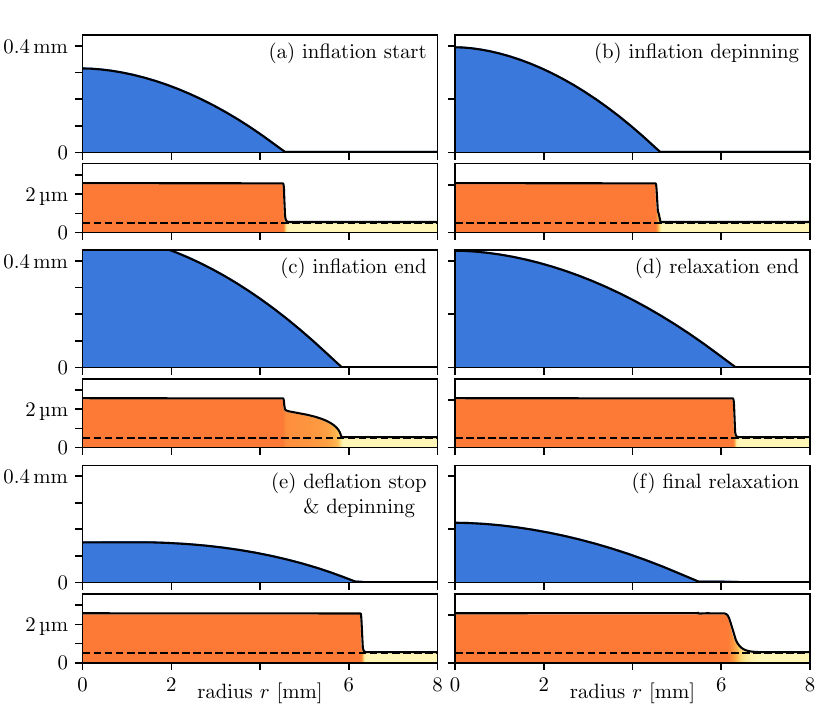}
  \caption{Snapshots of the simulation of a full inflation-deflation cycle for a \SI{12}{\micro l}--\SI{30}{\micro l} droplet for a constant volume rate \(|q|=\SI{10}{\micro l / s}\) at selected times. (a) State before inflation begins (after some initial swelling). (b) Moment of contact line depinning during inflation. (c) State at the end of the inflation before relaxation and (d) after \SI{1}{min} of relaxation. (e) Drop is deflated again and depins. (f) Contact line retracts during the final equilibration.
    The dependence on time is best seen in \cref{fig:brush-pumping_trace} (dashed lines). The properties of the liquid and the brush are identical to the ones used in \cref{fig:brush-spreading-snapshots} with the exception of $\thetaeq=\SI{10}{\degree}$.}
  \label{fig:brush-pumping_snapshots}
\end{figure}

\begin{figure}[!tbp]
  \centering
  \includegraphics{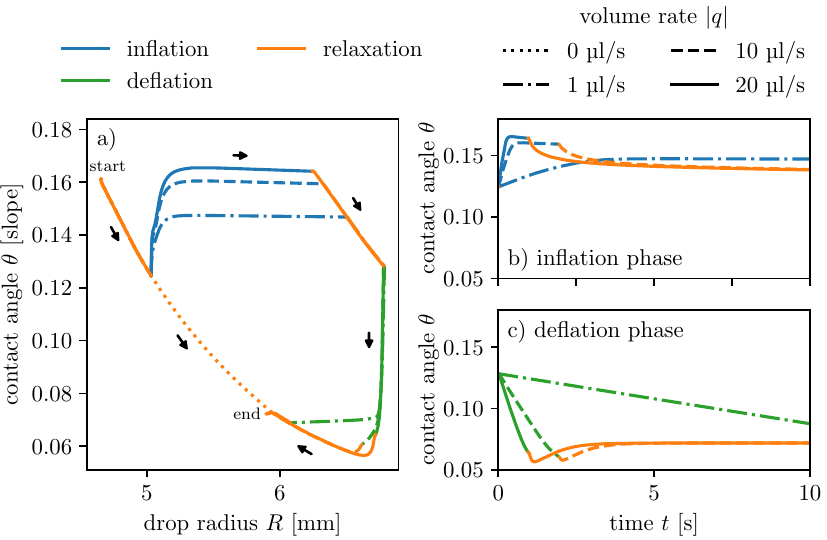}
  \caption{Contact angle measurements for inflation-deflation cycles of \SI{12}{\micro l} water droplets on a polymer brush at different flow rates \(|q|\). Panel (a) shows the contact angle against the drop base radius characterizing the drop shape and revealing the circular motion in the phase space. Panels (b) and (c) show the angle in dependence of the time since the beginning of the inflation/deflation process, respectively. The five phases are additionally distinguished by line color and the dash type indicates the magnitude of the volume rate. The properties of the liquid and the brush are identical to the ones used in \cref{fig:brush-spreading-snapshots} with the exception of $\thetaeq=\SI{10}{\degree}$.}
  \label{fig:brush-pumping_trace}
\end{figure}

As in the previous section, a small drop (here of \SI{12}{\micro l}) is initialized with the equilibrium shape on a dry brush. At \(t=0\) the film-brush interaction is ``switched on'' and the drop is allowed to relax on the swelling brush for a short time (here \SI{1}{min}). After the initial relaxation period, the droplet is inflated to a volume of \(\SI{30}{\micro l}\), followed by another relaxation phase. We limit the relaxation phases to \(\SI{1}{min}\) as a full equilibration of the brush would result in many hours of waiting time. After the drop relaxes in the inflated state, it is deflated back to its original volume, and the setup is left at \(q=0\) until it is fully equilibrated. Typical snapshots from such a cycle are given in \Cref{fig:brush-pumping_snapshots}. An effective representation of the behavior is given in \cref{fig:brush-pumping_trace} where panel~(a) gives the dependence of the contact angle $\theta$ on the drop radius $R$ (as frequently done in experimental work), while panels~(b) and (c) show the time-dependence of $\theta$ for the inflation and deflation phase, respectively. The results are given for three different pumping rates.

After the short initial relaxation phase where the drop spreads ($R$ increases and  $\theta$ decreases, upper left orange part in \cref{fig:brush-pumping_trace}~(a)), the drop is inflated. In this phase, at first the contact line remains pinned at its initial position ($R$ constant and  $\theta$ increases, nearly vertical start of blue part in \cref{fig:brush-pumping_trace}~(a)) until the contact angle passes a critical value and the droplet spreads  ($R$ increases and  $\theta$ constant, nearly horizontal start of blue part in \cref{fig:brush-pumping_trace}~(a)).
\Cref{fig:brush-pumping_snapshots}~(b) depicts the situation at the moment of depinning, and \Cref{fig:brush-pumping_snapshots}~(c) shows the readily inflated drop just before the relaxation phase. The contact line advances over the previously dry brush at a velocity slightly faster than the of of the ``swelling front.'' Consequently, at the end of the inflation phase the brush is not completely swollen below the entire drop. After waiting for \SI{1}{min}, the brush has swollen below the droplet ($R$ increases and  $\theta$ decreases, upper right orange part in \cref{fig:brush-pumping_trace}~(a)), but no significant swelling outside of the droplet has taken place (similar to the initial relaxation phase), as seen in \Cref{fig:brush-pumping_snapshots}~(d).

Early in the deflation phase, again the contact line remains pinned ($R$ constant and  $\theta$ decreases, green part in \cref{fig:brush-pumping_trace}~(a)). The hysteresis of the contact angle in the deflation stage (receding contact line) appears to be much larger than during the inflation stage (advancing contact line). As a result, depinning occurs relatively late during deflation, shortly before the drop reaches its original volume, cf.\ \Cref{fig:brush-pumping_snapshots}~(e).
The pinning at the edge of the swollen part of the brush is very pronounced. The final relaxation phase consists of two stages (lower right orange part in \cref{fig:brush-pumping_trace}~(a)). First, after depinning the drop relaxes fast (\cref{fig:brush-pumping_snapshots}~(f)) before on a much larger time scale the brush also swells in areas far from the drop. This leads to a small reduction in the drop volume, causing a subtle tail near the ``end'' label in \cref{fig:brush-pumping_trace}~(a). Unlike the case of a rigid substrate, the final drop shape differs from the initial state because the substrate has adapted, i.e.\ the process is not a closed \bfuwe{cycle.}

\Cref{fig:brush-pumping_trace} does not only show results for the single case of the inflation/deflation rate \(q\) used in \cref{fig:brush-pumping_snapshots} but also for a larger and a smaller rate.  We observe that the hysteresis between the advancing and receding dynamic contact angles is larger for larger \(|q|\). \bfuwe{Here, dedicated future experiments would allow one to assess whether certain measurement results for contact angle hysteresis actually depend on imposed flow rates, as mentioned in the review~\citenum{BLKS2022cocis}. Our results indicate that a proper static contact angle hysteresis should be determined in the limit of zero flow rate.}

However, the final state does not depend on \(q\) but is independent of the history as expected. Even a droplet that naturally spreads without any inflation (\(q=0\)), approaches the same equilibrium shape (see the dotted orange line in \cref{fig:brush-pumping_trace}). Simulations of repeated cycles show similar ``memory effects'' as found experimentally in Ref.~\citenum{SHNF2021acis} (not shown). \bfuwe{Further,  experimental results in Ref.~\citenum{FlMc2017l} indicate an increase in contact angle hysteresis with increasing adaptivity of the substrate and relate the observed more pronounced contact-line pinning to the more ``liquid-like behaviour'' of the contact region. Here, we report a similar effect, i.e.\ the difference between the dynamic advancing and receding contact angle significantly increases as compared to an inflation/deflation cycle on a rigid substrate at identical flow rate (not shown).  }

\begin{figure}[!tbp]
  \centering
  \includegraphics{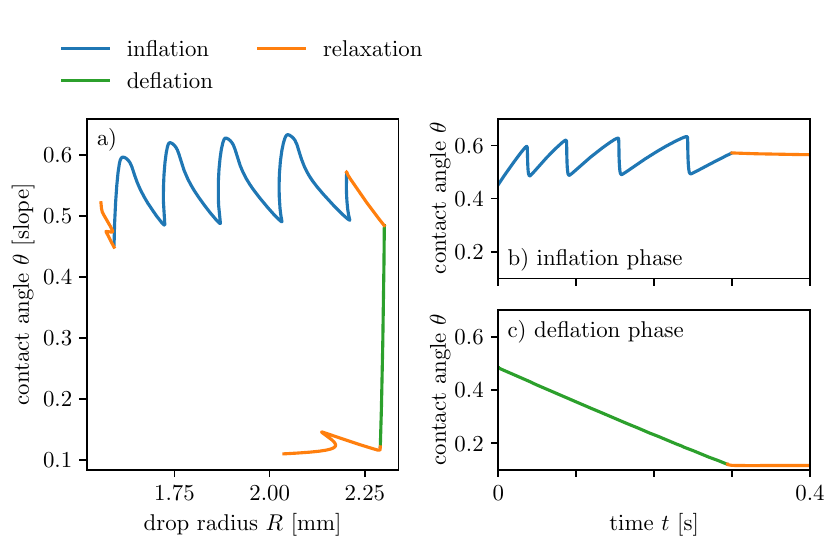}
  \caption{Contact angle measurements for inflation-deflation cycles as in \cref{fig:brush-pumping_trace}.
    The employed parameters differ from the ones used in \cref{fig:brush-pumping_snapshots} by an increased equilibrium contact angle ($\thetaeq=\SI{30}{\degree}$), an increased swelling rate (\(M_\text{im}=\SI{e-6}{m/Pa\,s}\)) and a weaker brush energy density due to \(\ell_K \approx \SI{20}{nm}\).
    Instead of the monotonic behavior of \cref{fig:brush-pumping_trace}, the inflation phase shows a pronounced stick-slip dynamics.}
  \label{fig:brush-pumping_stickslip}
\end{figure}

Although, the cycles studied in \cref{fig:brush-pumping_trace} all show some initial pinning in the phases of imposed in- and out-flux, both phases of forced contact line motion show monotonic behavior. However, the subtle interplay of the time scales of the various transport processes may also result in nonmonotonic behavior corresponding to pronounced stick-slip dynamics. This is illustrated in \cref{fig:brush-pumping_stickslip}.
There, a weaker brush energy density (\(\rho_\mathrm{brush} k_B T = \SI{500}{Pa}\), as \(\ell_K \approx \SI{20}{nm}\)) results in a more pronounced wetting ridge, which \bfuwe{develops faster} due to the increased swelling rate (\(M_\text{im}=\SI{e-6}{m/Pa\,s}\)).
In addition, the equilibrium (dry) contact angle parameter of the liquid has a value of \(\thetaeq=\SI{30}{\degree}\), all other parameters (besides $\ell_K$, $\thetaeq$ and $M_\text{im}$) are as in \cref{fig:brush-spreading-snapshots}. The larger wetting ridge causes a stronger initial sticking of the contact line as the droplet is inflated. The dynamic contact angle of the droplet steepens until it reaches a critical value, where the drop depins from the contact line, and then spreads in a rapid slipping motion.
Due to the fast swelling rate, a new wetting ridge can form as the liquid front slips, allowing for a repetition of the \bfuwe{stick and subsequent slip phase.}
Such a stick-slip motion \bfuwe{is} experimentally observed, for instance, in~Refs.~\citenum{WMYT2010scc,SHNF2021acis}. As in the experiments the stick-slip motion occurs in the inflation phase and manifests itself as sawtooth like structures in the \bfuwe{$(\theta,R)$-plane in Fig.~\ref{fig:brush-pumping_stickslip}~(a).} The deflation phase is, however, monotonic as before. Due to the radial symmetry the stick-slip motion is not perfectly periodic in time: The stick-phases get longer and the slip-phases cover more distance as the drop spreads. Although this well corresponds to the experimental results (e.g., Fig.~6 of Ref.~\citenum{WMYT2010scc}), the radial geometry does not easily lend itself to an analysis of the properties of stick-slip motion in dependence of the various influencing parameters. For such a study a planar geometry is preferable. Such a geometry, namely, the study of the advancing motion of a contact line on a brush-covered plate that is inserted at constant speed into a liquid bath~\cite{GrHT2023sm} is reviewed in the next section.

\subsection{Stick-slip in reverse Landau-Levich system}\label{sec:brush-stickslip}
Next, we investigate the behavior of a contact line that is forced to advance with a constant average velocity $U$. Experimentally this is realized, e.g., by pushing a brush-covered substrate into a liquid bath, i.e.\ via a reverse Landau-Levich geometry or by a doctor blade geometry, i.e.\ forcing a liquid film to advance over the substrate using a moving sharp blade.
As a simplistic but sufficient model, we consider a one-dimensional,\footnote{In contrast to the previous section, we do not consider a drop but liquid ridges or fronts in three dimensions that are translation-invariant in the direction transverse to the imposed velocity.}
long-wave two-field model, i.e.\ the model of Ref.~\citenum{ThHa2020epjt} corresponding to approximation~\ref{en:2field-brush-simp} above. Then, we perform a Galilean transformation from the reference frame of the brush-covered plate into the reference frame of the contact line region. In consequence, the first two equations of the model~\eqref{eq:brushevap-dynamic_eqs_final} are supplemented by the advective contributions $U\partial_x h$ and $U\partial_x \zeta$, respectively. The influence of the forcing is imposed by suitable boundary conditions, i.e.\ by allowing the brush to freely move through the domain boundary but restricting out-flux for the bulk liquid phase, cf.\ appendix~A.3 of Ref.~\citenum{GrHT2023sm}.

\begin{figure}[!tb]
  \centering
  \includegraphics{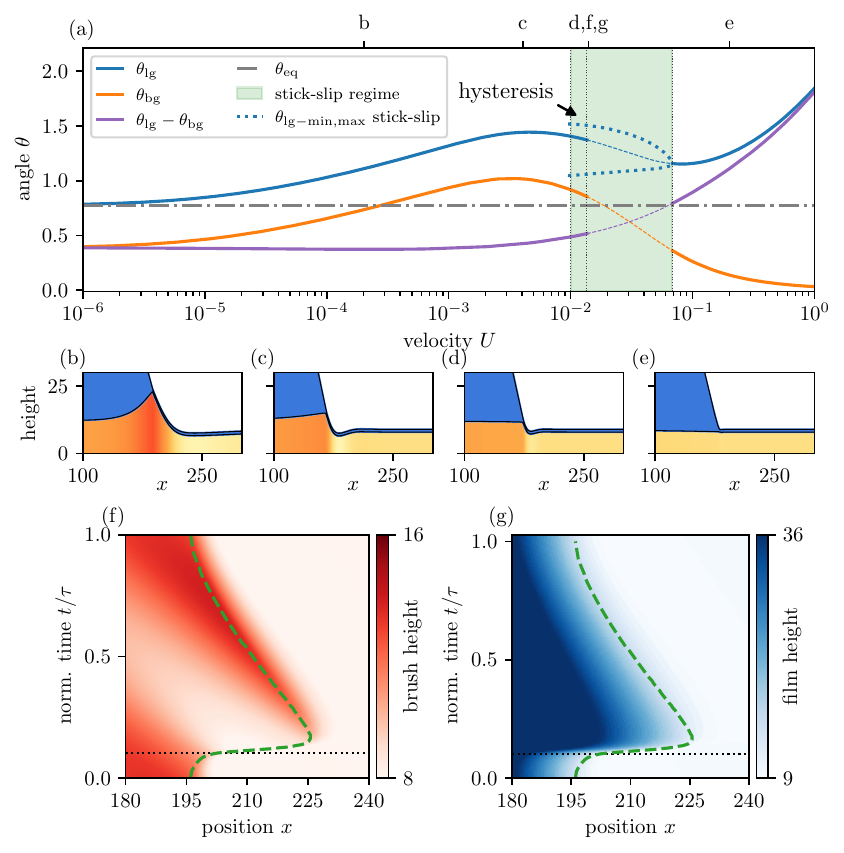}
  \caption{Panel (a) shows the dynamic angles $\theta_\mathrm{lg}$ and $\theta_\mathrm{bg}$ as well as their difference for a stationary state in dependence  of the forcing velocity $U$. Solid [dashed] lines indicate linearly stable [unstable] states. The velocity range, where time-periodic stick-slip behavior occurs is shaded in green. The blue dotted lines indicate the minimal and maximal contact angles during a stick-slip cycle.  Panels (b)--(e) show selected examples of the stationary profiles at the velocities indicated on the upper $x$-axis in panel~(a). Panels (f) and (g) show space-time plots of the brush and liquid height profiles, respectively, at $U=0.14$ for one stick-slip cycle of period $\tau=4800$. The green dashed line indicates the approximate position of the contact line, and the dotted horizontal line indicates when the contact line depins from the wetting ridge.  The long-wave approximation is used with parameters $T=0.02$, $\sigma=0.3$, $\gamma_\mathrm{bl}=0.3$, $\ell=20$, $M=0.1$ and $D=\SI{4e-3}{}$. \bfuwe{The figure summarizes selected data from figures 7, 8, 10 and 13 of Ref.~\citenum{GrHT2023sm} (used with permission of the Royal Society of Chemistry; permission conveyed through Copyright Clearance Center, Inc.).}} \label{fig:brush-forced_wetting}
\end{figure}

This setup allows us to investigate the contact line dynamics of forced wetting over several orders \bfuwe{of magnitude} of the forcing velocities $U$ as depicted in \cref{fig:brush-forced_wetting}. There, we summarize characteristics of the out-of-equilibrium configuration of the contact line\footnote{Here, we use nondimensional quantities. The velocity scale is $U_0 = \frac{3 h_\mathrm{a}^3 \eta \sqrt{\gamma_{\mathrm{lg}}}}{\sqrt{A_\mathrm{dry}}^3}$. Reintroducing the dimensions for an exemplary setting suggests that the stick-slip instability is triggered at $U \approx 0.55 \,\mathrm{mm}/\mathrm{s}$, cf.\ appendix~A.2 of Ref.~\citenum{GrHT2023sm}.}. Panel~(a) gives the dynamic contact angles $\theta_\mathrm{lg}
$ and $\theta_\mathrm{bg}$ of stationary profiles, (b)--(e) illustrate the contact line region for selected values of $U$ and panels (f) and (g) show space-time plots of the brush-liquid and liquid-gas interface, respectively, for an exemplary stick-slip cycle as observed in the green-shaded velocity range in~(a).

We start with a discussion of the stationary states.  At small velocities, all angles converge to their equilibrium values, i.e.\ the macroscopic contact angle $\theta_\mathrm{lg}$ is governed by Young's law and the mesoscopic angle $\theta_\mathrm{bg}$ by Neumann's law. In the regime up to $U\approx 2\times10^{-3}$ the angle $\theta_\mathrm{lg}$ increases while the height of the wetting ridge decreases. In this regime the Neumann law remains approximately valid, i.e.\ the contact line region is merely rotated. The is implied by the approximately constant difference  $\theta_\mathrm{lg}-\theta_\mathrm{bg}$ (horizontal purple line in \cref{fig:brush-forced_wetting}~(a)). Then, up to $U\approx 2\times 10^{-2}$, the onset of stick-slip motion is observed, accompanied by two observations: First, the purple line in \cref{fig:brush-forced_wetting}~(a) deviates from the horizontal, i.e.\ the angles at the wetting ridge of the stationary state no \bfuwe{longer fulfil} Neumann's law. Second, $\theta_\mathrm{lg}$ decreases with increasing velocity, i.e.\ $\partial  \theta_\mathrm{lg}/\partial U<0$. Varying the transfer rate $M$ reveals that the latter condition is necessary for the occurrence of stick-slip: As an increasing velocity favors a decreasing angle, the contact line has to advance even faster, which corresponds to a potentially destabilizing feedback loop. Finally, in the range of large velocities  $U>2\times10^{-2}$ the contact line moves so fast across the brush, that nearly no interaction can take place, and  the brush approximately behaves like a rigid substrate, see \cref{fig:brush-forced_wetting}~(e).  Comparing the stick-slip cycle in Figs.~\ref{fig:brush-forced_wetting}~(f) and (g) to the previously discussed behavior of an inflating droplet shows a similar sawtooth-like structure, i.e.\ at $t/\tau\approx 0.12$ (dotted horizontal line) the contact line rapidly depins from the wetting ridge and slips relatively fast until a new pinning wetting ridge starts to form. \bfuwe{A measurement of the contact angles during a stick-slip cycle shows, that depinning is related to the Gibbs criterion \cite{GASK2018prl,Quer2008armr,TDGR2014l,Gibb1874}. Namely, the contact line depins from the wetting ridge when the difference $\theta_\mathrm{lg}-\theta_\mathrm{bg}$, that corresponds to the contact angle of the liquid with respect to the ``dry'' descending flank of the wetting ridge,  exceeds the equilibrium contact angle.}

The combination of time simulations and stability analysis reveals that at low velocity the transition to stick-slip motion corresponds to a subcritical Hopf-bifurcation. Namely, there exists hysteresis, i.e.\ a velocity range  of bistability where the stick-slip dynamics and stationary state are both linearly stable. In contrast, the transition at high velocity is supercritical, i.e.\ no hysteresis is observed and the oscillation amplitude approaches \bfuwe{zero.} \bfuwe{The numerical analysis reveals, that the Gibbs criterion also plays a role for the latter transition. There, the  angle $\theta_\mathrm{lg}-\theta_\mathrm{bg}$ of the stationary flow state exceeds the equilibrium contact angle, i.e.\ the purple line in Fig.~\ref{fig:brush-forced_wetting}~(a) surpasses the horizontal grey dash-dotted line. Then, the contact line continuously slips and can no longer stick to a developing wetting ridge, which eliminates the stick-slip cycle.}

 Further analysis reveals, that stick-slip motion occurs in the velocity range, where the relevant time scales of the growth of the wetting ridge and of the forced motion are comparable. This observation ultimately allows for the prediction of the upper onset velocity as explained in detail in section~5.3 of Ref.~\citenum{GrHT2023sm}. \bfuwe{Dedicated experiments employing brush-covered plates in the here studied reverse Landau-Levich system would allow to reach a deeper understanding of onset and properties of stick-slip motion. In conjunction with theoretical work they could even be employed to experimentally determine system parameters that are difficult to access otherwise, e.g.\ relaxation rates and diffusion constants.}

\subsection{Sliding droplets}\label{sec:brush-sliding}
In the previous section we have investigated contact line dynamics within a reversed Landau-Levich geometry, i.e.\ a substrate is pushed into a liquid bath, e.g., at a certain inclination angle. Next, we consider drops of non-volatile liquids that slide down an inclined polymer brush-covered substrate under the influence of gravity. Thereby, we consider the drop size to be sufficiently small such that for a stationary state the influence of gravity on the drop shape can be neglected.
In the present section we use the full-curvature formulation and consider a one-dimensional substrate, therefore, the equilibrium drop profile corresponds to a circular arc. As in the case of the relaxational spreading dynamics, the considered evolution equations correspond to the two-field model of the limiting case~\ref{en:2field-brush}.

To account for \bfuwe{the force} that drives the drop down the incline, we add the corresponding part of the gravitational potential to the free energy, i.e.,
\begin{equation}
  \mathcal{F}=\widetilde{\mathcal{F}}+\int \beta (h+\zeta) x\, \mathrm{d}x,
\end{equation}
where $\widetilde{\mathcal{F}}$ denotes the original energy functional as defined in \cref{eq:brushevap-free-energy-functional} and $\beta = \varrho g \sin \vartheta$ characterizes the strength of the downhill force via the constant liquid mass density $\varrho$, the gravitational acceleration $g$, and the substrate inclination angle $\vartheta$, cf.~Ref.~\citenum{OrDB1997rmp}. The prefactor $\beta$ has the units of a force density.
Recomputing the variations of the free energy, we find that the potential simply adds to the conserved fluxes (cf.~Ref.~\citenum{GTLT2014prl})
\begin{equation}
  \vec j_h = -\frac{h^3}{3\eta} \, \left[ \nabla \frac{\delta \widetilde{\mathcal{F}}}{\delta h} + \beta \right]\quad \text{and} \quad
  \vec j_\zeta = -\frac{D_\mathrm{brush} \, \zeta}{\rho_\mathrm{liq} k_B T} \, \left[ \nabla \frac{\delta \widetilde{\mathcal{F}}}{\delta \zeta} + \beta \right].
\end{equation}
The transfer flux $j_\mathrm{im}$ remains unaffected as the contributions cancel. Furthermore, we expect that the additional contribution to the diffusive flux within the brush $\vec j_\mathrm{\zeta}$ is negligible.
Further, we only consider small inclination angles $\vartheta$, i.e.\ small $\beta$. Initially, the brush is assumed to be sufficiently dry.
Note that in the numerical simulations of the outlined system, for which we again use \texttt{oomph-lib}~\cite{HeHa2006}, we employ periodic boundary conditions, i.e, the same drop can move repeatedly through the simulation domain. In consequence, the simulation will never reach a stationary state in the laboratory frame. However, here, we only consider drops that do not cross the domain boundaries.

We use the total time derivative of the free energy to compute the free energy loss rate, namely, the dissipation:
\begin{align}
  \frac{\text{d}\mathcal{F}}{\text{d}t}= &\int\left[\frac{\delta \mathcal{F}}{\delta h}\partial_t h+\frac{\delta \mathcal{F}}{\delta \zeta}\partial_t\zeta\right]\text{d}x \nonumber
  =                                      \int \left[\frac{\delta \mathcal{F}}{\delta h}\left(-\partial_x j_h - j_\mathrm{im}\right)+\frac{\delta \mathcal{F}}{\delta \zeta}\left(-\partial_x j_\zeta+j_\mathrm{im}\right)\right]\text{d}x\nonumber\\
  =                       &-\int \frac{h^3}{3\eta}\left(\partial_x\frac{\delta \mathcal{F}}{\delta h}\right)^2\text{d}x-\int D\zeta\left(\partial_x\frac{\delta \mathcal{F}}{\delta\zeta}\right)^2\text{d}x-\int M_\mathrm{im}	\left(\frac{\delta \mathcal{F}}{\delta h}-\frac{\delta \mathcal{F}}{\delta \zeta}\right)^2\text{d}x\nonumber\\
  =:                                     &-D_h-D_\zeta-D_M\, \label{eq:Dissipation}
\end{align}
where $D_h$, $D_\zeta$, and $D_M$ represent the overall dissipation via viscous motion, diffusion, and liquid transfer, respectively. In the second step in~\eqref{eq:Dissipation} we have used partial integration (also cf.~Ref.~\citenum{GrHT2023sm}). The integrands of the corresponding integrals represent the local (per area) dissipation.

\begin{figure}[!tbp]
  \centering
  \includegraphics[width=\textwidth]{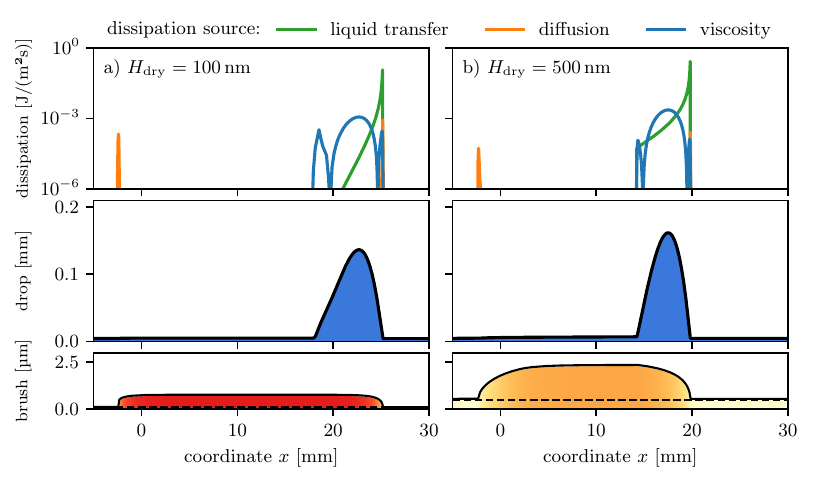}
  \includegraphics[width=0.8\textwidth]{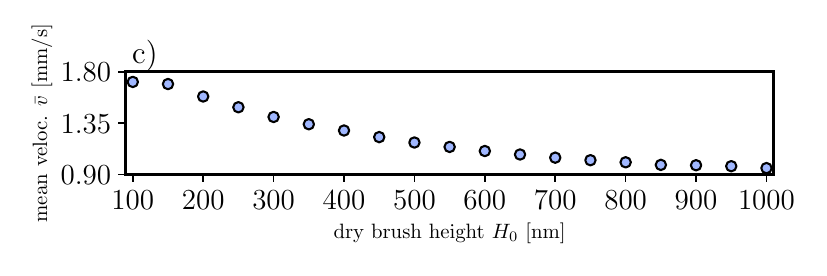}
  \caption{Snapshots of stationary sliding drops on inclined polymer brush-covered substrate for two different dry brush heights $H_\mathrm{dry}$ at identical inclination. In particular, in (a) a thin brush with $H_\mathrm{dry}=\SI{100}{nm}$ is considered while in (b) a thick brush with $H_\mathrm{dry}=\SI{500}{nm}$ is used. The simulations are initialized with a droplet of identical volume positioned on a dry brush at $x=\SI{0}{mm}$, and the snapshots are both taken after a time ($t=\SI{12}{s}$) has passed.
    The top panels show the local dissipation rates due to viscous flow (blue lines), liquid diffusion in the brush (orange lines) and liquid transfer (green lines). The central and bottom panels show the drop and brush profiles, respectively. Note the different vertical scales. \bfuwe{Panel~(c) presents the dependency of droplet speed on dry brush (for details see main text).} Here, we use $\thetaeq=\SI{10}{\degree}$, $\ha=\SI{4}{\micro m}$, $\beta=\SI{1e-5}{N/m^3}$. The remaining parameters are as in \cref{fig:brush-spreading-snapshots}. }
  \label{fig:brush-inclined_dissipation}
\end{figure}

\bfuwe{Panels (a) and (b) of \cref{fig:brush-inclined_dissipation}  present snapshots of stationary sliding drops on a brush-covered incline} for two different dry brush heights at identical inclination. Namely, on the left a thin brush with $H_\mathrm{dry}=100\,\mathrm{nm}$ is used while on the right a much thicker brush of $H_\mathrm{dry}=500\,\mathrm{nm}$ is used. Although each simulation is initialized with a droplet of identical volume at the same position, the drop on the thinner brush has moved farther, revealing that its average velocity is higher than the one of the drop on the thicker brush. This observation is in agreement with the fact that liquid droplets slide slower on softer substrates~\cite{CaSh2001l,MoAK2022el} as brushes with small [large] dry thickness $H_\mathrm{dry}$ can be interpreted as a rather rigid [soft] substrate. \bfuwe{\Cref{fig:brush-inclined_dissipation}~(c) presents the dependency of droplet speed on dry brush height.\footnote{\bfuwe{In particular, we determine the average center of mass speed for the time the drop slides between the positions $x=10\,\mathrm{mm}$ and $x=30\,\mathrm{mm}$ when it has well reached stationary sliding. For the considered cases the change in velocity due to liquid transfer into the brush is below one percent within the corresponding time window.}} We find that for the considered range of $H_\mathrm{dry}$ the speed monotonically decreases with increasing $H_\mathrm{dry}$. At small $H_\mathrm{dry}$ the speed seems to saturate indicating that the range of very thin brushes should in the future be studied more in detail. There, experiments indicate a nonmonotoneous behaviour \cite{Butt2023SPPworkshop}. We emphasize that, in principle, our dry brush height $H_\mathrm{dry}$ may be varied by changing either the grafting density $\sigma$ or the number of monomers per polymer chain. In \cref{fig:brush-inclined_dissipation} the increase in $H_\mathrm{dry}$ is due to an increase in the chain length, i.e.\ we keep $\sigma = 0.1$. }

Each column in \cref{fig:brush-inclined_dissipation} includes a top panel that visualizes the spatial distribution of the three components of dissipation. Note that the scale is logarithmic. This shows that the brush significantly brakes the droplet: While the viscous dissipation in the liquid bulk and at the contact line notably contribute to the free energy loss (blue curves), the dissipation also peaks at the imbibition front of the swollen brush due to the diffusive processes within the brush (orange curves). However, the largest contribution to dissipation is caused by the liquid transfer from drop to brush (green curve), i.e.\ by the swelling dynamics due to liquid transfer. Ultimately, this explains why the drop on the thicker brush is slower than the one on the thinner brush. Comparing the thin and the thick brush one notes that the transfer-induced dissipation is generally larger on the thicker brush. In both cases, it shows a sharp peak near the advancing contact line, i.e.\ where most of the swelling occurs. Although the peak is only slightly larger for the thicker brush, overall, the dissipation due to mass transfer-induced swelling is \SI{230}{\percent} larger for the thicker than for the thinner brush.
In contrast, for rigid solid substrates the dissipation is mostly due to viscous dissipation close to the contact line, as numerically demonstrated, e.g., in Ref.~\citenum{MoVS2011el} and Refs.~\citenum{EWGT2016prf,LBYM2023NC} for drops on one- and two-dimensional substrates, respectively. Here, the viscous dissipation close to the contact line remains relatively small as shear within the liquid is reduced by liquid transfer into the brush. This interpretation is supported by the observation that the viscous dissipation in the contact line region is actually smaller than the bulk contribution and is further diminished for the thicker brush.

\subsection{Spreading drops of weakly volatile liquid}\label{sec:drop-vol}

Finally, we consider the full three-field model given in its hydrodynamic form by Eqs.~\eqref{eq:brushevap-dynamic_eqs_final}, \eqref{eq:brushcons-fluxes} and \eqref{eq:brushevap-fluxes} and discuss corresponding results of numerical simulations originally published in Ref.~\citenum{KHHB2023jcp}. There, simulations of radially symmetric spreading drops of weakly volatile liquid on a polymer brush are quantitatively compared to matching experiments.

A typical simulation is initialized by placing a droplet on a dry brush in dry air. The drop shape corresponds to the equilibrium drop for a nonvolatile liquid on a rigid substrate with the wetting properties of a dry brush. The droplet has an initial volume of about \(\SI{0.3}{\micro l}\). At \(t=0\), evaporation and liquid-substrate transfer are switched on by setting $M_\mathrm{ev}$ and $M_\mathrm{im}$ to their intended values.

In the experiment, the spreading of a hexadecane droplet on a PLMA brush is considered in two different setups: (a) an ``open system'' where the drop and brush are exposed to the laboratory's atmosphere, i.e.\ the vapor particles can diffuse freely. Case (b) is a ``closed system'' where the drop and brush are confined in a small air-tight chamber of a low height \(d\). Case (b) can be directly described by our model, as it assumes a narrow-gap geometry for the vapor phase. The open case (a) we describe with our model by allowing the vapor particles to freely escape from the system at the lateral outer domain boundary, \(r=L\). In particular, we impose a constant (nonzero) humidity at \(r=L\), imitating contact with the lab atmosphere. Due to this Dirichlet boundary condition, the diffusive flux of vapor at \(r=L\) freely adapts. The closed case (b) is implemented by assuming zero diffusive flux at \(r=L\), i.e.\ by imposing a homogeneous Neumann condition for the vapor particle density \(\rho_\mathrm{vap}\).
The boundary conditions for the film and brush profile as well as their pressures are natural homogeneous Neumann conditions, suppress any flux of liquid through the domain boundaries.

\begin{figure}[!bt]
  \centering
  \includegraphics[width=\textwidth]{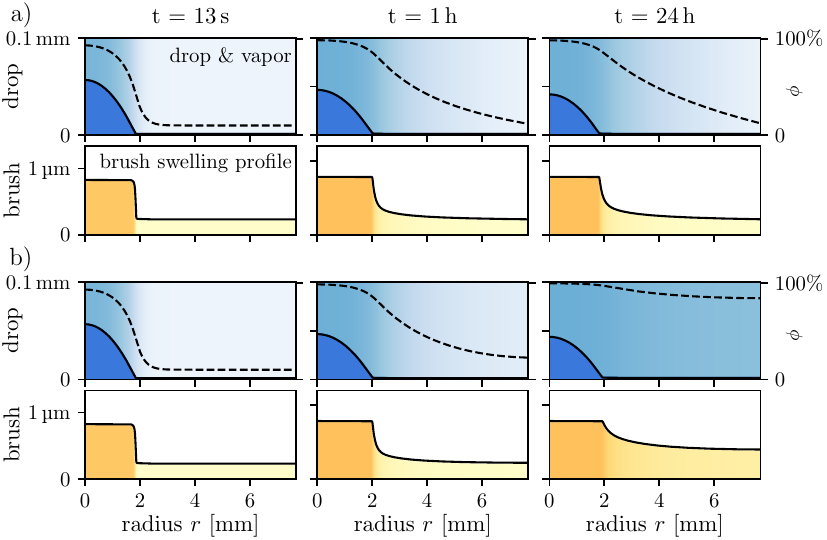}
  \caption{Snapshots from simulations of radially symmetric spreading drops of weakly volatile liquid on a polymer brush for (a) an open system where the vapor can escape at the outer boundary and (b) a closed system, i.e.\ all particles remain within the system. Respective upper panels give the vapor concentration \(\phi\) as light blue shading and dashed black lines and the drop profile in dark blue. The respective lower panel shows the brush profile as solid line and also indicates the swelling state by orange shading.
    The open and closed configurations behave differently at late times: in the open case (a) the brush develops a quasi-stationary halo while in the closed case (b) it continuously swells. The parameters are given in table~I of Ref.~\citenum{KHHB2023jcp}. Corresponding rates of important dynamic processes are depicted in \cref{fig:brushevap-fluxes}. \bfuwe{The figure was originally published as Fig.~6 of Ref.~\citenum{KHHB2023jcp} under the Creative Commons license CC BY.}}\label{fig:brushevap-snapshots}
\end{figure}

\begin{figure}[hbt!]
  \centering
  \includegraphics[width=\textwidth]{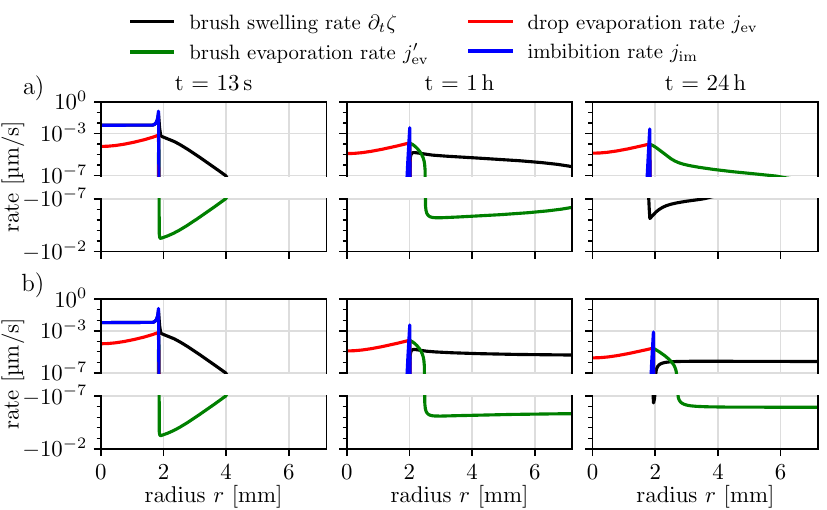}
  \caption{Rates of important dynamic processes corresponding to the (a) open and (b) closed case of \cref{fig:brushevap-snapshots}. In particular, we show the local swelling rate \(\partial_t \zeta\) of the brush (black) and the local transfer fluxes, namely, the drop-to-vapor evaporation rate \(j_\mathrm{ev}\) (red), the brush-to-vapor evaporation rate \(j_\mathrm{ev}'\) (blue), and the drop-to-brush imbibition rate \(j_\mathrm{im}\) (green), all as defined in \cref{eq:brushevap-fluxes}. Note that the contact line region is always situated at the left termination of the green line. The rates are measured on a symmetric logarithmic axis with a gap between the positive and negative parts. \bfuwe{The figure extends the data originally published in Fig.~7 of Ref.~\citenum{KHHB2023jcp} under the Creative Commons license CC BY.}}\label{fig:brushevap-fluxes}
\end{figure}

\Cref{fig:brushevap-snapshots} and \cref{fig:brushevap-fluxes} show typical \bfuwe{snapshots and corresponding fluxes,} respectively, at selected times for both, (a) the open and (b) the closed configuration. Since hexadecane is only weakly volatile and the vapor diffuses only slowly through the narrow gap, the droplet evaporates very slowly and is still present even after more than a \bfuwe{day.} For the snapshots, we have hence chosen three points in time, specifically, a few seconds, an hour, and one day after the initial placement of the drop.
In addition to the snapshots, we also provide deeper insights into the dynamic evolution of the system by plotting the local swelling rate and transfer fluxes for each snapshot in \cref{fig:brushevap-fluxes}.

Since the weakly volatile liquid takes some time to saturate the brush and the vapor, the frames at \(t=\SI{13}{s}\) and \(t=\SI{1}{h}\) are nearly identical for the open and the closed geometry: Within the first few seconds, the brush swells strongly directly underneath the droplet and the atmosphere absorbs vapor locally above the drop. Swelling and evaporation are strongest in the vicinity of the contact line (\cref{fig:brushevap-fluxes}). Some swelling also occurs beyond the drop's base and is driven by condensation of ambient vapor rather than by diffusive transport within the brush.
Note that both, the brush and the vapor phase, are in the diffusion-limited regime, i.e.\ the vertical transfer (including evaporation/condensation) is faster than lateral diffusion. This is also seen at \(t=\SI{1}{h}\). At this time brush and vapor in direct contact with the drop (\(r<R\)) have fully saturated while only a limited amount has diffused to the nearer periphery of the drop. This has created not only a gradient of vapor concentration in the atmosphere but also a gradient in the brush swelling around the drop. Here, an interesting interplay of the brush and vapor dynamics is observed: While the liquid within the brush in the direct vicinity of the droplet contributes to the evaporation into the vapor phase \(j_\mathrm{ev}'>0\), further away from the drop vapor condenses into the brush \(j_\mathrm{ev}'<0\). The strength of this effect strongly depends on the ratio of \bfuwe{diffusion in brush and vapor} (and on the sorption isotherm).

Dramatic differences between the open and closed case only become apparent in the long-time behavior: After several hours (see results for \(t=\SI{24}{h}\)) in the closed case (b) brush and atmosphere gradually saturate with liquid and vapor, respectively. This is well indicated by the significant swelling of the brush in areas that are far from the drop. In contrast, in the open case the swelling profile of the brush has nearly not changed. Instead a kind of stationary halo is established by ongoing nonequilibrium processes, namely, by balancing slow fluxes that overall are driven by the fixed low vapor concentration of the lab atmosphere \(\phi_\mathrm{lab}=\SI{10}{\percent}\) that is imposed at the boundary of the computational domain. Specifically, after \SI{24}{h} of time, in the open case (a) there is no longer any condensation of vapor particles into the brush. Instead, the system uses the transport within the brush layer as a way to transport liquid to areas far away from the drop, where \bfuwe{it evaporates into the less saturated vapor.} Thereby, the brush-to-vapor evaporation rate \(j_\mathrm{ev}'\) decreases with the distance from the drop and is generally orders of magnitude smaller than the evaporation rate observed in the direct vicinity of the drop. This indicates that the swelling of the brush far from the drop follows to some degree the local vapor concentration (or vice versa) via the sorption isotherm \(\phi(\alpha)\).
In other words, the brush profile could also be approximated in experiment or simulation by \bfuwe{determining} the vapor profile and converting it to a swelling profile using the sorption isotherm. We have checked this for the simulation in \cref{fig:brushevap-snapshots} and found that the results for the swelling profile obtained by time simulation and via sorption isotherm can not be distinguished by eye (not shown).

In contrast to the open case, in the closed case (b), the brush continuously swells in the areas far from the drop both, due to transport within the brush and due to condensation from the vapor. However, also here locally swelling profile and vapor phase are in equilibrium as described by the sorption isotherm (not shown).

To summarize, one striking difference to our simulations of a nonvolatile liquid on a polymer brush, specifically \cref{fig:brush-spreading-snapshots}, is that here in the open case the brush does not swell homogeneously in the long-time limit. This is due to the persistent gradient in the vapor concentration and the corresponding brush-to-vapor evaporation mechanism described above. In both, the experiments of Ref.~\citenum{KHHB2023jcp} and our simulations, the gradient of the brush swelling profile remains rather constant over time, i.e.\ the brush develops a quasi-stationary halo around the droplet that persists for the entire time until the drop has completely vanished (here, after several days).

\section{Conclusion}
\label{sec:conc}

In the present work, we have discussed recent advances in the hydrodynamic modeling of the dynamics of droplets on adaptive substrates. In particular, we have reviewed how the gradient dynamics form of mesoscopic hydrodynamics may be employed to extend the reach of thin-film (lubrication, full-curvature or long-wave) models toward the description of the dynamics of drops and films of volatile (and nonvolatile) liquids on brush-covered solid substrates. Mainly, we have summarized and expanded our recent corresponding work~\cite{ThHa2020epjt,GrHT2023sm,KHHB2023jcp}. We have as well discussed how a recently introduced mesoscopic model for volatile liquids that captures the coupled liquid and vapor dynamics in a narrow gap emerges as a limiting case~\cite{HDJT2023jfm}. After having introduced the gradient dynamics approach for films/drops of nonvolatile simple liquids on rigid inert solid substrates we have first expanded it to an arbitrary number of coupled \bfuwe{scalar degrees of freedom} before considering the specific case of drops of volatile liquids on brush-covered solids.

The resulting gradient dynamics model accounts for the coupled spreading, absorption, imbibition, diffusion, evaporation/condensation and swelling dynamics that occurs when a liquid drop is placed on a dry polymer brush. The underlying energy functional accounts for capillarity, wettability, brush energy, and entropic vapor contributions. In particular, wettability and capillarity (partly) depend on the brush state and adapt as the brush swells due to imbibition. After presenting the model in the general volatile case we have first discussed several limiting cases that correspond to models in the literature. Furthermore, we have illustrated its usage by considering the natural and forced spreading of drops of nonvolatile liquids as well as by discussing analyses of stick-slip motion and of sliding drops. Finally, also the spreading of volatile liquids has been considered. In passing we have discussed that Young's law remains valid for sessile droplets on the adaptive substrate, however, with a small correction due to the brush energy. Further, we have emphasized that the swelling brush may form a wetting ridge at the contact line. On the corresponding length scale a local Neumann law holds. We have also shown that the static spatially uniform limit of our model by be employed to discuss the sorption isotherm for a brush layer in contact with vapor.

The exemplary study of the relaxational spreading of a drop on the polymer brush has shown an intricate dynamics consisting of several qualitatively different phases in contrast to the spreading of a droplet on a rigid substrate. This is mostly due to the continuous adaptation of the substrate wettability, which acts as an additional driving force to the spreading motion. Our calculations have shown that the macroscopic Young angle as well as the Neumann angles relax exponentially towards their equilibrium values, in agreement with the exponential relaxation of the brush-fluid interface energies suggested by Ref.~\citenum{BBSV2018l}. For future studies, it would be interesting to investigate in detail how the behavior changes with varying ratios of the time scales related to the various involved processes. A comparison to dedicated experiments would also be welcome.

The study of the relaxational spreading has been complemented by a study of forced wetting and dewetting via the inflation and deflation, respectively, of a sessile drop. In this way we have replicated a typical experimental technique used to probe substrates covered by polymer brushes, hydrogels or other adaptive and/or flexible layers~\cite{WMYT2010scc,SHNF2021acis}. Unlike on a rigid substrate, the process is irreversible on the polymer brush, which is consistent with the experiments. Specifically, the contact angle of the liquid drop after the brush has been wet by liquid is lower than the contact angle before the forced wetting. In agreement with Ref.~\citenum{WMYT2010scc} for a range of brush properties and contact line speeds, the advancing contact line exhibits a stick-slip motion. Such behavior is also mentioned in Ref.~\citenum{SHNF2021acis} and is also described for other soft or otherwise deformable substrates~\cite{GCRB2004l,PGGS2007l,PuSe2008l,SAKL2011l,SKLL2011jcis,NCFM2012sm,KDNR2013sm,KBRD2014sm,KDGP2015nc,PBDJ2017sm,GASK2018prl,MoAK2022el}. This includes layers of thermoresponsive PNIPAM and related copolymers~\cite{GCRB2004l}. For such systems it might be interesting to elucidate the respective role of swelling and phase transition in the stick-slip behavior employing an amended gradient dynamics model. Other substrates coated with seemingly rigid nonabsorbing organic layers also show stick-slip behavior that is commonly related to heterogeneities and corresponding selective liquid condensation from the vapor phase~\cite{LWLH2002acis,TYYA2006l}, or is believed to result from behavior of chemical side groups of the employed polymer~\cite{SKCL2010jcis}. In view of the presented results on adaptive substrates it should be scrutinized whether even a small amount of swelling could also be an important factor for the observed stick-slip behavior. In general, stick-slip phenomena are widely occurring in the motion of contact lines of solutions and suspensions (with volatile solvent), see Ref.~\citenum{Thie2014acis} and references therein.

Subsequently, the stick-slip motion of a straight advancing contact line on a brush-covered substrate, i.e.\ in a planar geometry, has been considered reviewing Ref.~\citenum{GrHT2023sm}.  In particular, we have discussed a reverse (or inverse) Landau–Levich geometry where the brush-covered substrate is immersed into (instead of being drawn from) a liquid reservoir. This is similar to the use of a Wilhelmy plate technique described in Refs.~\citenum{PGGS2007l,PuSe2008l}. Here, we have numerically investigated in which range of parameters the stick-slip motion occurs, and have discussed how period and amplitude of the stick-slip cycle depend on substrate velocity. Further, we have related the onset conditions at low and high velocity to \bfuwe{Gibbs' criterion for depinning} and a cross-over in time scales, respectively. We expect that the model can be further refined after pertinent experimental results will become available. Further, detailed comparisons of the bifurcation character of the onset behavior as analyzed with the tools of dynamical systems theory between the present case and other stick-slip phenomena promises to deliver interesting results in the future.

The final presented example for nonvolatile liquids have been drops sliding down a brush-covered incline. After introducing the amended model that accounts for the downhill force due to gravity (or another body force) we have shown by time simulations how the presence of a brush results in a sliding velocity that decreases with increasing brush thickness. A preliminary analysis of the different dissipation channels has indicated that the effect is due to an effect similar to ``viscoelastic braking'' known from elastic substrates. There, it refers to situations where substrate dissipation dominates over liquid dissipation~\cite{CaSh2001l,PBDJ2017sm,MoAK2022el,HeST2021sm}. Here, however, the main contribution beside viscous dissipation within the liquid is the mass transfer between liquid drop and brush layer that is the driver of the deformation of the brush, e.g., of the advancement of a swelling front (related to the entropic elasticity of the brush).

Finally, the spreading of a drop of volatile liquid has been briefly discussed, mainly referring to results of Ref.~\citenum{KHHB2023jcp}. The shown example has revealed how local evaporation and condensation dynamics interact with transport in vapor phase and brush to stabilize a nearly stationary inhomogeneous nonequilibrium swelling profile -- a so-called halo -- that occurs in an open geometry but not in a closed one. A quantitative comparison of experiments and simulations is shown in Ref.~\citenum{KHHB2023jcp} while here we have only discussed the usage of the model. \bfuwe{Remarkably, the model may be used to provide estimates for constants that are difficult to measure experimentally, e.g., the diffusion constant of oil within a brush.}
The reviewed results emphasize that transport through the brush can be crucial. The final example has also illustrated that an incorporation of the long-wave description of the coupled dynamics of vapor and liquid phase developed in Ref.~\citenum{HDJT2023jfm} into more complicated mesoscopic hydrodynamic models is facilitated by the gradient dynamics approach.

However, in several aspects the presented model is still rather basic and allows for several ways toward future improvements: We have assumed that within the brush transport of liquid only occurs by diffusion. In consequence, there is no dynamic coupling between motion in the brush and the liquid in the drop. As a result, only the components on the main diagonal of the mobility matrix for the conserved dynamics are nonzero. Treating the liquid flow within the brush layer in more detail will be cumbersome but feasible. For instance, one may consider the brush to have flow properties similar to a porous layer and use methods developed for the coupling of flow on and in a thin porous layer (cf.~\citenum{ThGV2009pf} and references therein). However, based on Ref.~\citenum{ThGV2009pf} it is to expect that main features can already be captured by a brush state-dependent effective slip at the brush-liquid interface.

Furthermore, the versatility of the gradient dynamics approach would allow one to expand the reviewed cases toward more complex situations. For example, to consider an adaptive substrate that interacts with a liquid mixture would be of high interest. Such systems have recently attracted much attention in the context of the co-nonsolvency transition~\cite{BKTW2017l, Somm2017m, YREU2018m, YBUF2019m, SHNF2021acis}. To model drops of mixtures on brushes, one would need to combine the approach presented here with gradient dynamics models for films of mixtures, see, e.g., Ref.~\citenum{ThTL2013prl}. Also insoluble and soluble surfactants may be accounted for~\citenum{ThAP2012pf,ThAP2016prf}. In general, it should be possible to describe most of the adaptive substrates discussed by Ref.~\citenum{BBSV2018l} with gradient dynamics models by amending and extending the examples we have presented. \bfuwe{Also the application of the modeling approach to slippery lubricant-infused surfaces (SLIPS), lubricant-infused surfaces (LIS), slippery covalently attached liquid surfaces (SCALS) should be discussed \cite{BLKS2022cocis,GLNK2023acie,GrNe2023acis,CHRT2023nrc}.}

\section*{Acknowledgements}
This work was supported by the Deutsche Forschungsgemeinschaft (DFG) within SPP~2171 by Grants No.\ TH781/12-1 and TH781/12-2. We acknowledge fruitful discussions on adaptive viscous and viscoelastic substrates with Christopher Henkel and Jacco Snoeijer, as well discussions on the interaction of liquids and polymer brushes with the group of Sissi de Beer. Furthermore we acknowledge a Strategic Collaboration Grant by the University of Twente and the University of M\"unster (2019-2020).


\providecommand{\latin}[1]{#1}
\makeatletter
\providecommand{\doi}
  {\begingroup\let\do\@makeother\dospecials
  \catcode`\{=1 \catcode`\}=2 \doi@aux}
\providecommand{\doi@aux}[1]{\endgroup\texttt{#1}}
\makeatother
\providecommand*\mcitethebibliography{\thebibliography}
\csname @ifundefined\endcsname{endmcitethebibliography}
  {\let\endmcitethebibliography\endthebibliography}{}

\end{document}